\documentclass[apj,twocolumn,numberedappendix]{openjournal}
\usepackage{amsmath}
\usepackage{booktabs}
\usepackage{multirow}
\usepackage{color}
\usepackage{soul}
\usepackage{booktabs} 
\usepackage{adjustbox}

\usepackage[dvipsnames]{xcolor} 

\usepackage[breaklinks,colorlinks,citecolor=blue,urlcolor=blue]{hyperref}

\defcitealias{Shen2018}{S18}

\renewcommand{\arraystretch}{1.2}  
\usepackage{listings}
\usepackage{color}
\definecolor{dkgreen}{rgb}{0,0.6,0}
\definecolor{gray}{rgb}{0.5,0.5,0.5}
\definecolor{mauve}{rgb}{0.58,0,0.82}
\definecolor{golden}{rgb}{0.86,0.65,0.01}
\usepackage{orcidlink}

\begin{document}


\title{A generative model for {\it Gaia} astrometric orbit catalogs: selection functions for binary stars, giant planets, and compact object companions}

\author{Kareem El-Badry\,\orcidlink{0000-0002-6871-1752}$^{1,2}$}
\author{Casey Lam\,\orcidlink{0000-0002-6406-1924}$^{3}$}
\author{Berry Holl\,\orcidlink{0000-0001-6220-3266}$^{4,5}$}
\author{Jean-Louis Halbwachs\,\orcidlink{0000-0003-2968-6395}$^{6}$}
\author{Hans-Walter Rix\,\orcidlink{0000-0003-4996-9069}$^{2}$}
\author{Tsevi Mazeh\,\orcidlink{0000-0002-3569-3391}$^{7}$}
\author{Sahar Shahaf\,\orcidlink{0000-0001-9298-8068}$^{8}$}

\affiliation{$^1$Department of Astronomy, California Institute of Technology, 1200 E. California Blvd., Pasadena, CA 91125, USA}
\affiliation{$^2$Max-Planck Institute for Astronomy, K\"onigstuhl 17, D-69117 Heidelberg, Germany}
\affiliation{$^3$Observatories of the Carnegie Institution for Science, 813 Santa Barbara St., Pasadena, CA 91101, USA}
\affiliation{$^4$Department of Astronomy, University of Geneva, Chemin Pegasi 51, 1290 Versoix, Switzerland}
\affiliation{$^5$Department of Astronomy, University of Geneva, Chemin d'Ecogia 16, 1290 Versoix, Switzerland}
\affiliation{$^6$Université de Strasbourg, CNRS, Observatoire astronomique de Strasbourg, UMR 7550, 11 rue de l’Université, Strasbourg, France}
\affiliation{$^7$School of Physics and Astronomy, Tel Aviv University, Tel Aviv, 6997801, Israel}
\affiliation{$^8$Department of Particle Physics and Astrophysics, Weizmann Institute of Science, Rehovot 7610001, Israel}

\email{Corresponding author: kelbadry@caltech.edu}

\begin{abstract}
Astrometry from {\it Gaia} DR3 has produced a sample of $\sim$170,000 Keplerian orbital solutions, with many more anticipated in the next few years. These data have enormous potential to constrain the population of binary stars, giant planets, and compact objects in the Solar neighborhood. But in order to use the published orbit catalogs for statistical inference, it is necessary to understand their selection function: what is the probability that a binary with a given set of properties ends up in a catalog? We show that such a selection function for the {\it Gaia} DR3 astrometric binary catalog can be forward-modeled from the {\it Gaia} scanning law, including individual 1D astrometric measurements, the fitting of a cascade of astrometric models, and quality cuts applied in post-processing. We populate a synthetic Milky Way model with binary stars and generate a mock catalog of astrometric orbits. The mock catalog is quite similar to the DR3 astrometric binary sample, suggesting that our selection function is a sensible approximation of reality. Our fitting also produces a sample of spurious astrometric orbits similar to those found in DR3; these are mainly the result of scan angle-dependent astrometric biases in marginally resolved wide binaries. We show that {\it Gaia's} sensitivity to astrometric binaries falls off rapidly at high eccentricities, but only weakly at high inclinations. We predict that DR4 will yield $\sim 1$ million astrometric orbits, mostly for bright ($G \lesssim 15$) systems with long periods ($P_{\rm orb} \gtrsim 1000$ d). We provide code to simulate and fit realistic  {\it Gaia} epoch astrometry for any data release and determine whether any hypothetical binary would receive a cataloged orbital solution. 
\end{abstract}

\maketitle

\section{Introduction}
\label{sec:intro}
The third data release of the {\it Gaia} mission \citep[DR3;][]{GaiaCollaboration2016, GaiaCollaboration2023} included astrometric orbital solutions for $\sim 168,000$ binary systems, representing a vast improvement in sample size and completeness over all previous work \citep{arenou_2023}. These data have already enabled discovery of several astrophysically interesting objects, including a sample of compact objects in au-scale orbits with stellar companions \citep[e.g.][]{Shahaf2023} and a sample of giant planets orbiting nearby stars \citep{Holl2023}. For a review of the {\it Gaia} binary star sample, see \citet{El-Badry2024}. 

Epoch-level astrometric data was not published in DR3, and several quality cuts were imposed on the published orbital solutions. These cuts depend on a variety of quantities, such as the signal-to-noise ratio of the photocenter semi-major axis, the orbital period, and the parallax and eccentricity uncertainties. These cuts -- combined with the fact that {\it Gaia's} sensitivity to astrometric orbits depends on quantities such as the eccentricity and orientation in ways that are difficult to predict analytically -- make it a nontrivial problem to use the {\it Gaia} astrometric binary sample for population interference, and few attempts at modeling the sample have been made so far.

In this paper, we use a model of the solar neighborhood's binary population to build a forward-model for the {\it Gaia} DR3 astrometric binary catalog. This entails modeling the {\it Gaia} observations at the level of individual epochs, predicting the observation times, scan angles, and simulated one-dimensional astrometry for each simulated source from the {\it Gaia} scanning law. We include common false positives and reproduce the cascade of single-star, accelerating, and full orbital astrometric models employed in producing the observed catalog. Because our modeling results in a mock catalog of orbital solutions that resembles the one actually published in DR3, we believe that it captures the most important selection effects and false positives of the real catalog. It can thus be used both to interpret the DR3 binary sample and to forecast what will be discoverable in future data releases. 

The remainder of this paper is organized as follows. In Section~\ref{sec:how_observe}, we summarize the basics of {\it Gaia} observations and the astrometric signal caused by a binary. Section~\ref{sec:galaxia} describes the Galactic model, simulated binary population, and the assumptions we make to predict 1D epoch astrometry. We discuss the astrometric model cascade in Section~\ref{sec:cascade} and present the resulting mock catalog in Section~\ref{sec:mock_catalog}. Finally, Section~\ref{sec:monte_carlo} presents a Monte Carlo selection function that can be used to model and fit epoch astrometry for arbitrary binaries. Section~\ref{sec:disc} summarizes our main results. 

This paper aims to develop a relatively detailed model of {\it Gaia} orbit catalogs by forward-modeling epoch astrometry and the astrometric model cascade. A companion paper, \citet{Casey}, develops a model that is more approximate but significantly less computationally expensive.

\section{How Gaia observes}
\label{sec:how_observe}
How {\it Gaia} observes has been summarized extensively in other work, and we refer to \citet{GaiaCollaboration2016} for a detailed description. We summarize the most important aspects here.

{\it Gaia} is equipped with two telescopes whose fields of view are separated by 106.5 degrees. The satellite rotates with a period of 6 hours, such that the two telescopes sweep out an annulus with a width of about 0.7 deg on the sky. The rotation axis precesses with a period of 63 days and a fixed tilt angle of 45 degrees with respect to the Sun direction, causing this annulus to rotate on the sky. At the same time, the spacecraft orbits the Sun with a period of a year. The combination of these three rotations results in full-sky coverage, though with the observing cadence and distribution of scan angles varying significantly across the sky \citep[for details, see][]{Holl2023b}. For stars brighter than $G=15$, the median number of visibility periods used in DR3 is 20, with a (16-84)\% range of 16-27. Here a visibility period refers to a group of observations separated by other groups of observations by at least 4 days. Data contributing to DR3 solutions were obtained over a period of about 1000 d.

As a source moves across the astrometric field of view (FOV), it is independently observed by 8-9 different CCDs. Each CCD observation results in an independent measurement of the source's position in the along-scan (AL) direction relative to a reference position assigned to the source at a reference time. Crucially, {\it only } a 1D measurement in the AL direction is made with high precision and used in the astrometric solution. At $G\lesssim 14$ -- which is the regime relevant to most astrometric binaries published so far -- the per-CCD observations currently reach an AL precision of order 0.12 mas. All the CCD transits in a given FOV transit are spread over a period of less than a minute, which is much shorter than the orbital periods of astrophysically plausible binaries to which {\it Gaia} is sensitive. The 8-9 CCD transits can thus be regarded as essentially independent measurements of the same quantity, and can be averaged to obtain smaller uncertainties in the CCD-averaged measurements. Our modeling in this work suggest that these measurements are indeed nearly independent and not systematics-dominated, such that the per-FOV transit uncertainties are $\sim 3$ times smaller than the per-CCD uncertainties (Appendix~\ref{sec:appendix_singlestar}).

{\it Gaia} astrometry is built upon a global astrometric solution, whose calculation requires the simultaneous optimization of millions of attitude and calibration parameters, and astrometric parameters for $\sim 100$ million stars \citep{Lindegren2012}. Most of these parameters are of little interest for the analysis of one source. It is possible to transform the astrometric measurements for a single source to ``local plane coordinates'' in a tangent plane to the unit sphere in the vicinity of each source, as described by \citet{Lindegren_local_plane}. The astrometric measurements required to describe the motion of a source can then be represented with just a small file containing the AL displacements, their uncertainties, and associated metadata (scan angles, parallax factors, transit times, etc.) for each source. Thus far, such epoch astrometry has been published for only one source: the binary Gaia BH3 \citep{Panuzzo2024}.

\section{Galactic model and binary population}

\label{sec:galaxia}
We now describe the Galactic model and binary population from which we forward-model {\it Gaia} observations and construct a mock astrometric orbit catalog. This will allow us to validate our selection function by comparing the mock catalog to the DR3 data.

\subsection{Galactic model}
Our modeling uses \texttt{Galaxia} \citep{Sharma2011}, which generates synthetic resolved-star surveys from the Besançon model of the Milky Way \citep{Robin2003}. We use a modified version of the code described by \citet{Lam2020}. We only attempt to model sources within 2 kpc of the Sun, because 99\% of all astrometric orbits published in DR3 are found within 2 kpc.

\texttt{Galaxia} predicts a total of 1.1 billion sources (all representing single stars) within 2 kpc of the Sun. Comparing to the {\it Gaia} DR3 \texttt{gaia\_source} catalog, we find that \texttt{Galaxia} predicts 1.4 times  more sources than are observed (with \texttt{parallax\_over\_error > 10}) within both 50 and 100 pc. Since {\it Gaia} is expected to be nearly complete within these volumes for sources above the hydrogen burning limit \citep{GaiaCollaboration2021_gcns}, we conclude that \texttt{Galaxia} predicts 40\% too many sources in the Solar neighborhood. We thus discard a random subset of predicted sources, chosen independent of distance and apparent magnitude, to reduce source counts by a factor of 1.4

\subsection{Binary population}
\texttt{Galaxia} does not include binary stars. We thus use \texttt{COSMIC} \citep{Breivik2020} to generate a zero-age binary population according to the model from \citet{Moe2017}. We assume a \citet{Kroupa2001} primary mass function\footnote{Note that this is different from the more top-heavy mass function assumed in COSMIC by default.} and draw mass ratios, periods, and eccentricities from the covariant, mass-dependent distributions inferred by \citet{Moe2017}. The assumed binary fraction is $\approx 41\%$ for solar-type primaries,\footnote{This is lower than $f_{{\rm mult}}\approx0.5$, the mean number of companions per solar-type primary, because higher-order multiples contribute more than one companion. Only companions with $0.2<\log\left(P_{{\rm orb}}/{\rm d}\right)<8$ are considered.} increasing to 57\% at $M_1=3\,M_{\odot}$ and 80\% at $M_1=6\,M_{\odot}$. Below $0.8\,M_{\odot}$, the binary fraction is assumed to decrease linearly with $\log M_1$, from 40\% at $0.8\,M_\odot$ to 0 at $0.08\,M_\odot$. Almost all of the binaries predicted to be detectable in DR3 have primary masses between 0.7 and $1.3\,M_{\odot}$, so it is the binary population in this mass range -- which is constrained primarily by the \citet{Raghavan2010} survey of nearby G stars -- that is most important for our modeling. 
 
\texttt{COSMIC} predicts populations of single and binary stars. We match both binaries and singles to the \texttt{Galaxia}-predicted sources, generating sources until the total number of stellar systems (singles plus binaries, with each binary representing a single system) in the zero-age \texttt{COSMIC}-predicted population exceeds the number of \texttt{Galaxia} sources within 2 kpc by 10\%. This accounts for the fact that the \texttt{Galaxia} population only contains stars that are still alive; for a Kroupa IMF and constant star formation history over 12 Gyr, about $\approx 10\%$ of all stars that have been born will have already died.  

Massive stars are preferentially found in the Galactic disk and thus are subject to higher extinction than typical lower-mass stars. To account for this, we place each simulated binary at the position of the  \texttt{Galaxia} star whose mass is closest to that of the primary. We assign inclinations by drawing from a $\sin(i)$ distribution. We draw longitudes of the ascending node, $\Omega$, and arguments of periastron, $\omega$, from $\mathcal{U}(0, 2\pi)$. We define the reference phase as $\phi_{0}=\frac{2\pi T_{p}}{P_{\rm orb}}$, where $T_p$ is the epoch of periastron, and draw $\phi_0$ from $\mathcal{U}(0,1)$.

We assign $G-$band absolute magnitudes to the primary and secondary components using the \texttt{isochrones} package \citep{Morton2015}, using \texttt{MIST} models \citep{Choi2016} and assuming a uniform age distribution between 0 and 12 Gyr. We remove binaries with initial masses $>8\,M_{\odot}$ whose primaries have died, assuming that most would-be black hole or neutron star binaries are destroyed or dramatically tightened by common envelope evolution and supernova kicks. We similarly assume that most binaries containing white dwarfs (WDs; i.e., primaries with initial masses $<8\,M_{\odot}$ that have terminated their evolution) shrink to short periods and are not detectable astrometrically. However, motivated by the results of \citet{Shahaf2024} and \citet{Yamaguchi2024}, we assume that 10\% of WD + luminous star binaries with initial separations of $(2-6)$\,au form wide post-common envelope binaries with separations that are 50\% of their initial separations. This prescription is quite simplified and is not expected to capture all aspects of the observed population. The 10\% fraction is chosen to approximately match the number of binaries with large mass functions in the observed catalog (Section~\ref{sec:mock}). We assign WD masses using the initial-final mass relation of \citet{Weidemann2000}. We remove binaries in which either star currently fills its Roche lobe or would have filled it previously (e.g., red clump stars that would have filled their Roche lobes at the tip of the first giant branch). Because most binaries that are astrometrically detectable contain main-sequence stars in au-scale orbits, the treatment of evolutionary effects has minor effects on the overall properties of the observable population. 

We calculate extinctions to each binary using the \texttt{combined19} dust map in the \texttt{mwdust} package \citep{Bovy2016}, which combines the maps from \citet{Drimmel2003}, \citet{Marshall2006}, and \citet{Green2019}. We assume $A_G = 2.8 E(B-V)$, where $E(B-V)$ is on the \citet{Schlegel1998} scale. We then calculate each binary's total $G-$band apparent magnitude, angular separation, $\rho$, and magnitude difference, $\Delta G = G_2-G_1$. Binaries that are resolved or marginally resolved are removed from the sample. As we show in Appendix~\ref{sec:appendix_ipd}, the transition between resolved and unresolved binaries depends on both $\rho$ and $\Delta G$, and in DR3 can be approximated as $ \Delta G\,\left[{\rm mag}\right]=\frac{1}{25}\left(\frac{\rho}{{\rm mas}}-200\right)$, where binaries with $\Delta G$ larger than this value are unresolved. We remove resolved binaries from the sample, as well as unresolved binaries with $G > 19$, which were not fit with binary solutions in DR3. We only generate mock astrometry for binaries, discarding single stars and neglecting both higher-order multiples and chance alignments of physically unassociated stars. We are left with 46 million unresolved binaries with $G < 19$ within 2 kpc.

As a consistency check, we calculated the total number of predicted binaries with absolute magnitude $M_{G}=3-6$ (since solar-type binaries dominate the astrometric binary sample in DR3), finding $9.2\times 10^4$ and $1.03 \times 10^6$ within 200 pc and 500 pc, respectively. These fractions represent 44\% and 45\%, respectively, of the total number of {\it Gaia} sources with $M_{G}=3-6$ within the same volumes. The good agreement between these fractions and the 41\% assumed binary fraction for solar-type stars suggests that the \texttt{Galaxia} model is reasonably accurate.

\begin{figure}
    \centering
    \includegraphics[width=\columnwidth]{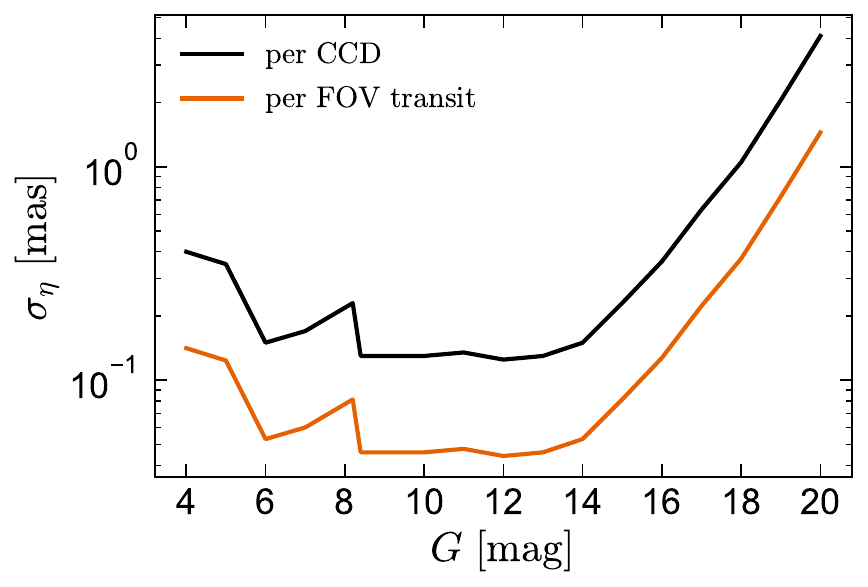}
    \caption{Assumed AL displacement uncertainties. Black line shows the uncertainty per CCD transit, adopted from \citet{Holl2023}. Tan line shows our assumed uncertainty per FOV transit and is smaller than the black line by a factor of $\sqrt{8}$, representing the result of averaging measurements from an average of 8 uncorrelated CCD measurements per FOV transit. The uncertainties are smallest at $G=8-14$ and are dominated by calibration systematics in this magnitude range, but we find (Appendix~\ref{sec:appendix_singlestar}) that measurements from individual CCDs can still be usefully combined to yield FOV transit-averaged uncertainties a factor of $\approx \sqrt{8}$ smaller. The uncertainties at $G \gtrsim 14$ are photon-limited and thus rise rapidly toward faint magnitudes. Calibration problems worsen at $G\lesssim8$. }
    \label{fig:noise}
\end{figure}

\subsection{Epoch astrometry}
We mock-observe each binary using the scan times and scan angles predicted by the {\it Gaia} Observation Forecast Tool (\texttt{GOST})\footnote{https://gaia.esac.esa.int/gost/}. Because querying \texttt{GOST} involves a relatively slow web query, we first pre-computed the scanning law at 49152 uniformly spaced points on the sky (healpix level 6), and then used the saved value at the point closest to each simulated source. This results in an effective resolution of about 0.9 degrees, which is comparable to the 0.7 degree across-scan FOV. 

When modeling astrometry for {\it Gaia} DR3, we only include observations between JD 2456891 and JD 2457902 \citep{Halbwachs2023}. For DR4 observations, we include scans taken between JD 2456891 and JD 2458868. Due to a variety of issues \citep[see][]{Lindegren2021}, about 10\% of FOV transits do not result in usable data. To account for this, we reject FOV transits for each source with 10\% probability.

\subsubsection{Noise model}
\label{sec:noise_model}
We add Gaussian noise to the predicted AL displacements according to the empirical median CCD AL abscissa uncertainty in DR3 \citep[][their Figure 3, ``EDR3 adjusted'' model]{Holl2023}. At $G\lesssim 13$, these uncertainties are on average a factor of $\approx 2$ larger than the formal uncertainties calculated by the image parameter determination (IPD) pipeline \citep[e.g.][]{Lindegren2021b}. This discrepancy is a result of imperfect treatment of systematics for bright sources. In the astrometric non-single star (NSS) pipeline, epoch uncertainties were correspondingly inflated, with the intention that the residuals for typical good solutions should be consistent with the uncertainties \citep[see][]{Holl2023}. While this uncertainty inflation was only partially successful (i.e., some trends in goodness-of-fit statistics with apparent magnitude still exist), it is important to remember that different uncertainties were assumed in calculating binary and single-star astrometric solutions. This implies that their goodness-of-fit metrics should not be directly compared, and that the two kinds of solutions are likely affected by different types of systematics \citep[e.g.][]{Nagarajan2024b}.

Each FOV transit contains 8 or 9 individual CCD transits, each separated by a few seconds. We bin these into a single CCD-averaged value per transit, with uncertainty $\sqrt{N_{\rm bin}}$ times smaller than the per-CCD uncertainty. Here $N_{\rm bin}$ represents the number of measurements from individual CCDs that are being averaged; we adopt $N_{\rm bin}=8$ to account for the fact that some CCD measurements flagged as outliers are discarded. We show the assumed per-CCD and per-FOV transit uncertainties in Figure~\ref{fig:noise}. We verified that applying this method of single sources fit with 5-parameter astrometric solutions results in astrometric uncertainties that are in good agreement with those published in {\it Gaia} DR3 (see Appendix~\ref{sec:appendix_singlestar}).

For sources brighter than $G=13$, we add unmodeled Gaussian noise to account for underestimated uncertainties due to various systematics, which have been shown to increase at $G<13$ \citep[e.g.][]{Lindegren2021, Lindegren2021b, El-Badry2021}. A sharp increase in \texttt{goodness\_of\_fit} of the DR3 astrometric binary solutions at $G<13$ \citep[e.g.][]{El-Badry2023_bh2} shows that this effect was not fully corrected by error inflation applied to the epoch astrometry in the NSS astrometric pipeline. We draw the per-FOV transit ``$\sigma$'' of the unmodeled noise from $\mathcal{U}(0,0.04)$ mas. This is motivated by our fit to the epoch astrometry for Gaia BH3 \citep{Panuzzo2024}, where we find that adding $0.04$ mas in quadrature to the per-FOV transit uncertainties results in a reduced $\chi^2$ of 1. 

\subsection{Generating mock observations}
\label{sec:mock}

We now consider how an unresolved or marginally resolved binary affects the epoch astrometry. Our modeling  follows \citet{Lindegren2022}. In a given scan, a source is observed with scan angle $\psi$, defined as degrees east of north. We define $\rho$ as the sky-projected, instantaneous angular separation of the two stars, and $\theta$ as the binary's position angle. Then  $\Delta \eta = \rho \cos(\psi-\theta)$ is the angular separation projected onto the scan angle. Finally, we define $\xi = \Delta \eta /u$, where $u$ quantifies the resolution of the {\it Gaia} line spread function. In detail, $u$ will depend on the flux ratio and other quantities, but \citet{Lindegren2022} and \citet{Holl2023b} find that $u=90$ mas provides a reasonable approximation for modeling astrometry of binaries, and we adopt this value throughout.

We assume that the two stars each contribute flux along the AL direction as 1D Gaussians with unit variance separated by $\xi$, and with amplitudes scaled according to the flux ratio, $f=10^{\left(G_{1}-G_{2}\right)/2.5}$, where $G_1$ and $G_2$ are the magnitudes of the primary and secondary. We further assume that the measured AL displacement will be the {\it peak} of the resulting flux profile. At sufficiently large angular separations, the two peaks will be resolved, and in this case we assume the IPD pipeline will correctly centroid the peak of the primary. We predict the AL displacement relative to the binary's center of mass in these two cases as

\begin{equation}
    \label{eq:bias}
    \delta \eta = 
\begin{cases} 
u B \left( f, \xi \right) - \frac{q}{1+q} \Delta \eta & \text{if } 0 < \left| \xi \right| \leq 3 - f, \\
- \frac{q}{1+q} \Delta \eta & \text{if } 3 - f < \left| \xi \right|.
\end{cases}
\end{equation}
Here $B(f,\xi)$ is obtained by iteratively solving the equation 
\begin{equation}
    \label{eq:B}
    B = \frac{f \xi }{f + \exp(\xi^2 / 2 - \xi B)}.
\end{equation}
for $B$, beginning at $B=0$. This yields the AL coordinate of the peak of the total flux profile (see \citealt{Lindegren2022}, for details). Inspection of Equation~\ref{eq:bias} reveals that in the limit of small $\xi$ (i.e., close separations), it reduces to $\delta \eta=\left(\frac{f}{f+1}-\frac{q}{1+q}\right)\Delta\eta$. In this case, the observations exactly trace the photocenter. As $\xi$ increases, the peak of the flux profile no longer exactly traces the photocenter, but is displaced toward the primary. For typical flux ratios, the difference between the first case in Equation~\ref{eq:AL} and the photocenter displacement becomes significant ($\gtrsim 0.05$\,mas) only for separations $\Delta \eta \gtrsim 30$\,mas. This means that {\it Gaia} effectively observes the photocenter for the vast majority of binaries with periods short enough to be astrometrically constrained. On the other hand, for marginally resolved wide binaries, {\it Gaia} does {\it not} observe the photocenter. The difference between the observed AL displacements and the photocenter will vary with scan angle, and this can lead to spurious orbits with a range of periods related to the scanning law, as we explore in Section~\ref{sec:spurious} and Appendix~\ref{sec:appendix_resolved}.


The 2nd case of Equation~\ref{eq:bias} corresponds to cases where two peaks are detected in the AL flux profile and the IPD processing correctly identifies the centroid of the primary. It is rarely relevant for binaries published in DR3, because resolvable pairs are removed by the cut on \texttt{ipd\_frac\_multi\_peak} used in selecting sources for processing with orbital solutions (Appendix~\ref{sec:appendix_ipd}). 

For widely separated pairs with $\xi > 0.5$, we assume that the marginally-resolved nature of the source leads to poor centroiding. We thus add unmodeled Gaussian noise with $\sigma=0.5$ mas to the epoch astrometry in epochs where $\xi > 0.5$. This choice is somewhat ad-hoc but is motivated by the fact that the model otherwise predicts too many spurious solutions with good formal fits, leading to features in the recovered period distribution that are not present in the DR3 data.

\begin{figure}
    \centering
    \includegraphics[width=\columnwidth]{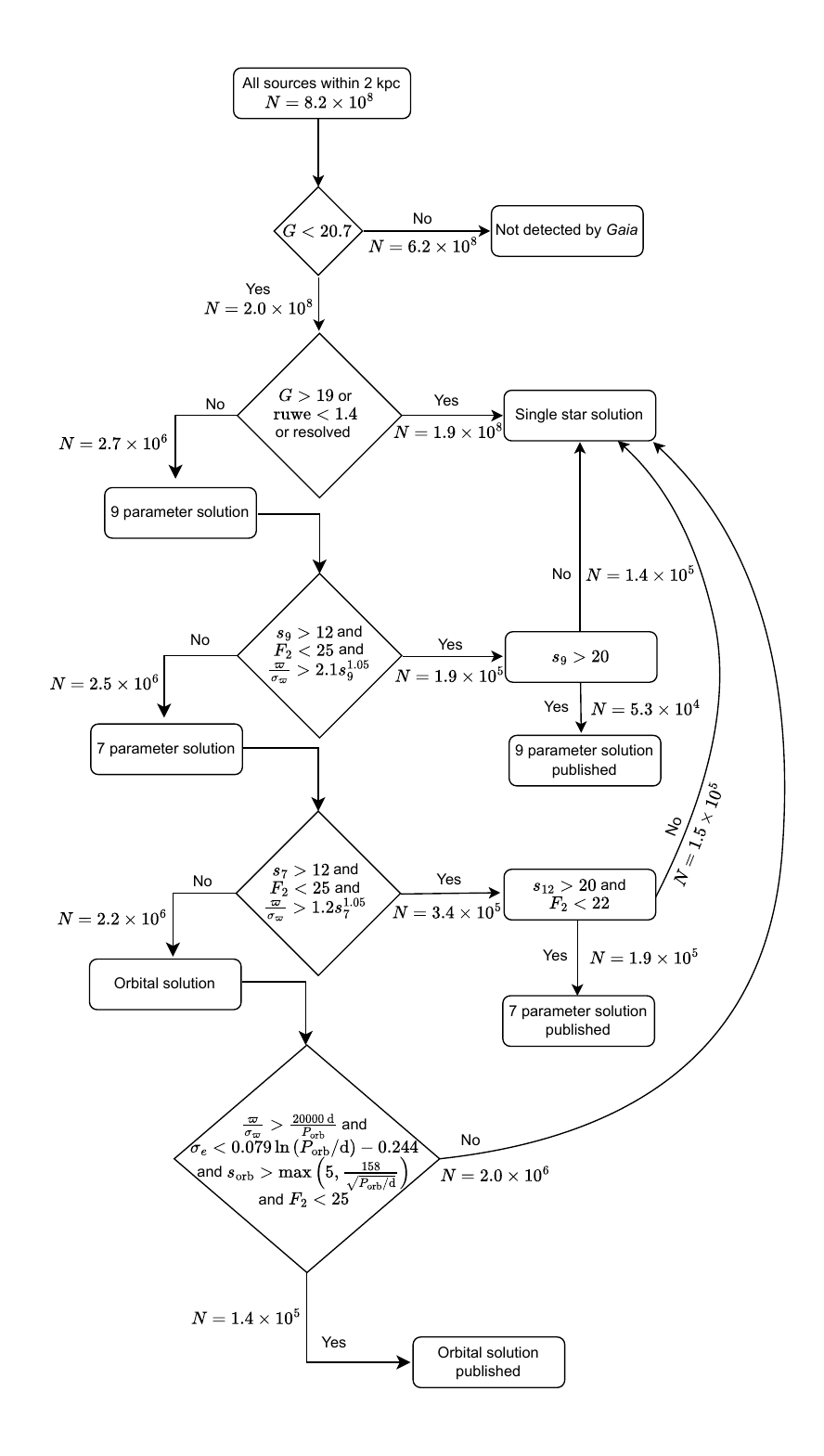}
    \caption{Astrometric model cascade used in constructing the mock catalog of orbital solutions. The cascade approximates the one used in generating the DR3 binary catalog \citet[][their Figure 1]{Halbwachs2023}. The guiding principle is to only fit more complex models (e.g. orbits) in cases where simpler models (single-star and acceleration solutions) produce a manifestly poor fit. The number of sources retained and excluded at each branch is indicated.Many of the orbital solutions rejected at the last step of filtering are spurious. Only a small fraction ($< 1\%$) of all binaries detected by {\it Gaia} within 2 kpc ultimately receive astrometric orbital solutions in DR3.}
    \label{fig:flowchart}
\end{figure}

\section{Fitting the astrometric data}
\label{sec:cascade}
We now describe how we fit the epoch astrometry generated above. We attempt to reproduce the procedure used to generate the binary solutions published in DR3 whenever possible, as described by \citet{Halbwachs2023}. The astrometric model cascade is illustrated in Figure~\ref{fig:flowchart}. We emphasize that our main goal is to reproduce the DR3 catalog of orbital solutions (i.e., \texttt{nss\_solution\_type = Orbital} or \texttt{AstroSpectroSB1}), so the cascade shown in Figure~\ref{fig:flowchart} is somewhat simplified compared to the one actually used in DR3. 

It is not known a priori which sources are single stars and which are detectable binaries or higher-order multiples. In {\it Gaia} DR3, the guiding principle was to publish the simplest astrometric solution that could  satisfactorily explain the data, beginning with a single-star model.

\subsection{5-parameter single-star solution}
All sources are first fit with a 5-parameter, single-star astrometric model. It predicts AL displacements given by 

\begin{equation}
    \label{eq:eta_5par}   \eta=\left[\Delta\alpha^{*}+\mu_{\alpha}^{*}t\right]\sin\psi+\left[\Delta\delta+\mu_{\delta}t\right]\cos\psi+\Pi_{\eta}\varpi.
\end{equation}

Here $\Delta\alpha^{*}$ and $\Delta \delta$ represent the source position at a reference time $t_{\rm ref}$, relative to the reference point $(\alpha_0, \delta_0)$ assigned to each source.\footnote{For {\it Gaia} DR3, $t_{\rm ref}=\rm J2016.0=JD\,2457389.0$. For DR4, we anticipate $t_{\rm ref}=\rm J2017.5=JD\,2457936.875$.} $\mu_\alpha^*$ and $\mu_\delta$ are proper motions, and $\varpi$ is the source parallax. 
The scan angles, $\psi$, and parallax factors, $\Pi_\eta$, are precomputed for each source by \texttt{GOST}; the latter are defined such that multiplying the true parallax by $\Pi_\eta$ gives the source's parallactic motion in the AL direction at the time of the relevant scan. Note that $\psi$ and $\Pi_\eta$ are functions of time.

Given a list of observation times, $t_i$, measured AL displacements, $\eta _i$, their uncertainties, $\sigma_{{\eta,i}}$, the scan angles $\psi_i$, and parallax factors, $\Pi_{\eta,i}$, the maximum-likelihood values of the astrometric parameters $\left\{\Delta_\alpha^*, \mu_\alpha^*, \Delta \delta, \mu_\delta, \varpi \right\}$ and their uncertainties can be obtained via linear regression. These parameters can then be used to calculate the predicted AL displacements, $\eta_{{\rm pred},i}$, and a corresponding $\chi^2$ statistic: 
\begin{equation}
    \label{eq:chi}
    \chi^{2}=\sum_{i}\frac{\left(\eta_{{\rm pred,}i}-\eta_{i}\right)^{2}}{\sigma_{\eta,i}^{2}}.
\end{equation}

At this stage it is necessary to correct for the fact that we average (``bin'') measurements from different CCDs within a FOV transit, reducing the number of astrometric measurements by a factor of $N_{\rm bin} = 8$. In general, neither $\chi^{2}$ nor reduced $\chi^{2}$ is conserved under binning. Assuming that all CCDs yield independent measurements of the same quantity, we derive an expression for the expected $\chi^{2}$ of the unbinned data in terms of the $\chi^{2}$ of the binned data:
\begin{equation}
    \label{eq:binned_chi2}
    \chi_{{\rm unbinned}}^{2}=\chi_{{\rm binned}}^{2}+N_{{\rm FOV\,transits}}\left(N_{{\rm bin}}-1\right).
\end{equation}
In the subsequent text, $\chi^2$ always refers to the value calculated for the unbinned data via Equation~\ref{eq:binned_chi2}.

From this we calculate a ``unit weight error'', 
\begin{equation}
    \label{eq:uwe}
    {\rm UWE}=\sqrt{\frac{\chi^{2}}{\nu}},
\end{equation}
where $\nu$ is the number of degrees of freedom. In this case, $\nu=N_{{\rm FOV\,transits}}N_{{\rm bin}}-5$; i.e., the number of unbinned datapoints minus the number of free parameters. UWE is analogous to the \texttt{RUWE} statistic published in {\it Gaia} DR3 \citep{Lindegren_2018_RUWE, Lindegren2021}, where the latter is renormalized to correct for empirical trends in the astrometric uncertainties with color and apparent magnitude. It would {\it not} be appropriate to renormalize our calculated UWE to match the {\it Gaia} \texttt{RUWE}, because the latter is calculated from underestimated epoch-level uncertainties while the former is calculated from uncertainties that have already been inflated to yield UWE $\sim 1$ for single sources (Section~\ref{sec:noise_model}). We verify that our UWE values are a good approximation of {\it Gaia's} \texttt{RUWE} in Section~\ref{sec:ruwe}.

Various authors have shown that many sources with \texttt{RUWE} $> 1.4$ are binaries \citep[e.g.][]{Belokurov2020, Penoyre2022}. However, not all binaries are expected to have \texttt{RUWE} $> 1.4$, since in some cases the orbital period is too long or the astrometric uncertainties are too large for the companion to produce detectable deviations from single-object motion. In {\it Gaia} DR3, only sources with  \texttt{RUWE} $> 1.4$ were processed with binary models. We thus fit all 46 million simulated unresolved binaries satisfying $G<19$ with single-star solutions, discarding those with UWE $< 1.4$. Following the DR3 procedure, we also discarded $3\times 10^4$ sources that were observed in fewer than 12 visibility periods. 
 
\subsection{Acceleration solutions}
\label{sec:accel}
The simplest extension of the single-star solution is a 7-parameter acceleration solution, which adds two free parameters, $\dot{\mu}_\alpha^*$ and $\dot{\mu}_\delta$, for acceleration in the right ascension and declination directions: 

\begin{equation}
\begin{aligned}
 \eta = & \left[ \Delta\alpha^{*} +  \mu_{\alpha}^{*}t + \frac{1}{2} \dot{\mu}_\alpha^*t^2 \right] \sin \psi \\
& + \left[ \Delta\delta +  \mu_\delta t + \frac{1}{2} \dot{\mu}_\delta t^2 \right] \cos \psi 
 + \Pi_\eta \varpi.
\end{aligned}
\label{eq:7par}
\end{equation}
This model is appropriate if the deviation from single-object motion can be satisfactorily described by a constant accelerations, as is expected for binaries with orbital periods much longer than the observing baseline.

A similar 9-parameter model can be defined for sources with variable acceleration, as might be expected for binaries with periods only a few times longer than the observing baseline:

\begin{equation}
\begin{aligned}
 \eta = & \left[ \Delta\alpha^{*} +  \mu_{\alpha}^{*}t + \frac{1}{2} \dot{\mu}_\alpha^*t^2 + \frac{1}{6} \ddot{\mu}_\alpha^*t^3 \right] \sin \psi \\
& + \left[ \Delta\delta +  \mu_\delta t + \frac{1}{2} \dot{\mu}_\delta t^2 + \frac{1}{6} \ddot{\mu}_\delta t^3 \right] \cos \psi 
 + \Pi_\eta \varpi.
\end{aligned}
\label{eq:9par}
\end{equation}
Here $\ddot{\mu}_\alpha^*$ and $\ddot{\mu}_\delta$ represent the acceleration derivatives.\footnote{Equations~\ref{eq:7par} and ~\ref{eq:9par} are not identical to the models fit by \citet[][their Equations 6 and 7]{Halbwachs2023} because they include an additional time offset to make the source positions at the reference epoch similar between accelerating and single-star solutions. This does not affect the magnitude of the inferred accelerations, the significance, or the goodness-of-fit.}

Adding additional free parameters will in general always lead to a better fit, so it is useful to quantify whether the data significantly constrain the additional free parameters. For acceleration solutions, ``significance'' is quantified as the modulus of the two-dimensional vector of additional free parameters (relative to the next-simplest model) divided by its uncertainty: 

\begin{figure}
    \centering
    \includegraphics[width=\columnwidth]{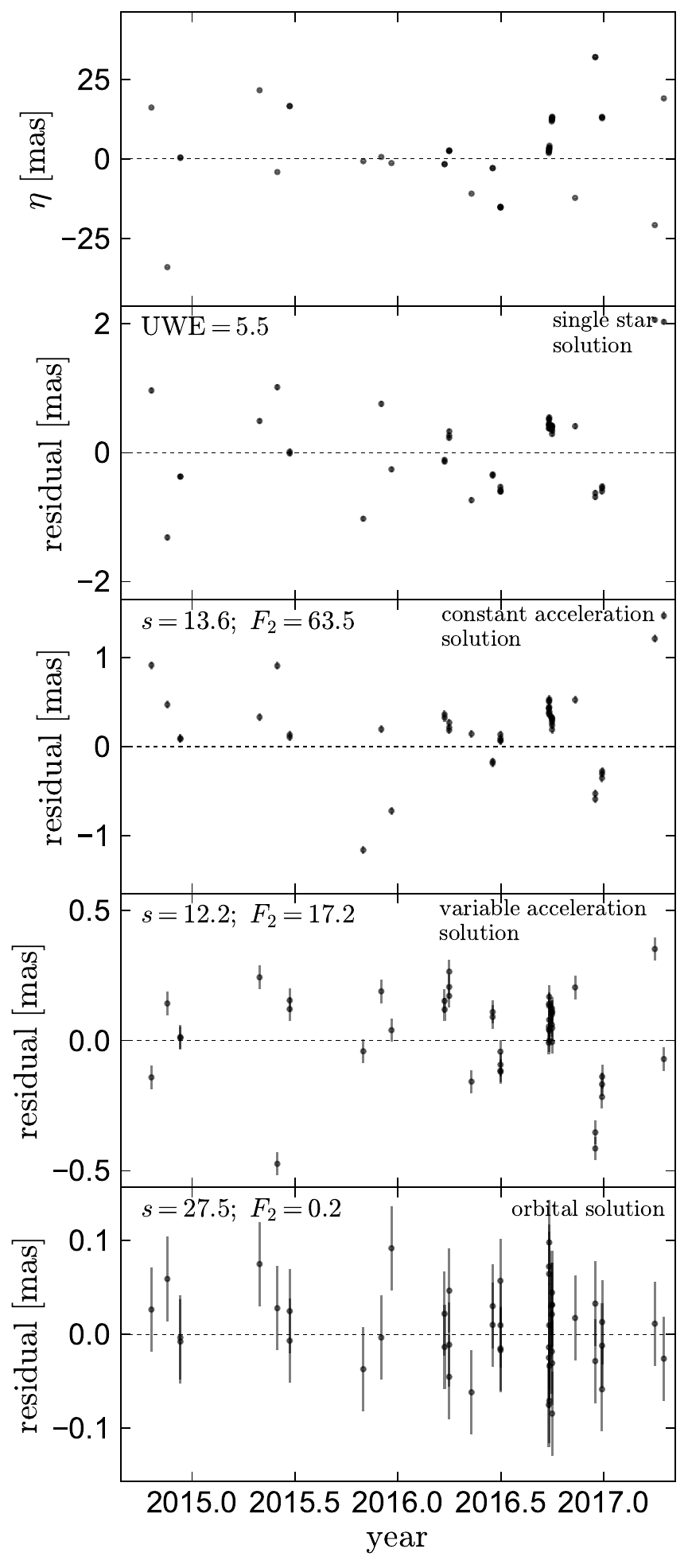}
    \caption{Example epoch astrometry for a simulated binary with $P_{\rm orb} = 902$\,d, $e=0.5$, and $\varpi=2.5$\,\rm mas. Top panel shows the raw AL displacements. Second panel shows residuals with respect to the best-fit single star model, which provides a manifestly bad fit, as quantified by UWE = 5.5. Third and 4th panels show residuals for 7- and 9-parameter acceleration solutions. Although neither of these yields a very good solution, the 9-parameter (variable acceleration) solution provides a good enough fit ($F_2 < 25$ and $s > 12$) that it would have been provisionally accepted in the processing of DR3, and thus no orbital solution would have been attempted for this source. Bottom panel shows the residuals with respect to the best-fit Keplerian orbital solution, which yields a good solution that would have been accepted {\it if} the source reached that stage in the astrometric solution cascade. }
    \label{fig:epoch_astrometry}
\end{figure}

\begin{equation}
    \label{eq:sig}
    s = \frac{1}{\sigma_1 \sigma_2} \sqrt{\frac{p_1^2 \sigma_2^2 + p_2^2 \sigma_1^2 - 2 p_1 p_2 \rho_{12} \sigma_1 \sigma_2}{1 - \rho_{12}^2}}.
\end{equation}
Here $p_1$ and $p_2$ represent the additional parameters of the model, $\sigma_1$ and $\sigma_2$ their uncertainties, and $\rho_{12}$ the correlation coefficient between them. For the 7-parameter model (Equation~\ref{eq:7par}), $p_1$ and $p_2$ represent the acceleration terms, $\dot{\mu}_{\alpha}^*$ and  $\dot{\mu}_{\delta}$; for the 9-parameter model, they represent the acceleration derivatives,  $\ddot{\mu}_{\alpha}^*$ and  $\ddot{\mu}_{\delta}$.

The goodness-of-fit of acceleration and orbital solution is quantified by the $F_2$ statistic \citep{Wilson1931}, which is a transformation of $\chi^2$:

\begin{equation}
    \label{eq:F2}
    F_2 = \sqrt{\frac{9 \nu}{2}} \left[ \left( \frac{\chi^2}{\nu} \right)^{1/3} + \frac{2}{9 \nu} - 1 \right].
\end{equation}
For well-behaved solutions with reliably-estimated uncertainties, $F_2$ is expected to follow a normal distribution, $\mathcal{N}(0,1)$. Values of $F_2$ significantly above 1 suggest a poor solution or underestimated uncertainties. To correct for underestimated uncertainties, the uncertainties of the astrometric parameters are re-scaled by a constant,  
\begin{equation}
    c=\sqrt{\frac{\chi^{2}}{\nu\left(1-\frac{2}{9\nu}\right)^{3}}},
\end{equation}
such that sources with poor fits have inflated uncertainties at fixed apparent magnitude.

In {\it Gaia} DR3, the 9-parameter variable acceleration solution (Equation~\ref{eq:9par}; {\it not} the simpler 7-parameter model) was the first model tried for sources with \texttt{RUWE} $>1.4$. the acceptance criteria applied before testing the orbital model were:

\begin{equation}
\label{eq:accept}
\begin{cases}
s_9 > 12 \\
F_2 < 25 \\
\frac{\varpi}{\sigma_\varpi} > 2.1 s_9^{1.05}
\end{cases}
\end{equation}
where $s_9$ is the significance of the 9-parameter solution (Equation~\ref{eq:sig}) and $\varpi$ and $\sigma_\varpi$ are the parallax and parallax uncertainty of the same solution. 

Solutions satisfying these cuts were provisionally accepted with 9-parameter solutions (i.e., other models were not tried, but only solutions passing additional cuts were actually published). Sources not passing Equation~\ref{eq:accept} were fit with 7-parameter acceleration solutions and (provisionally) accepted  using similar criteria:
\begin{equation}
\label{eq:accept_7}
\begin{cases}
s_7 > 12 \\
F_2 < 25 \\
\frac{\varpi}{\sigma_\varpi} > 1.2 s_7^{1.05}
\end{cases}
\end{equation}

As described by \citet{Halbwachs2023}, the acceleration solutions selected at this stage do not constitute the full sample published in the \texttt{nss\_acceleration\_astro} catalog, as some lower-significance ``alternative' solutions were added at later stages and more stringent cuts on significance and $F_2$ were applied later in filtering (Figure~\ref{fig:flowchart}). But importantly for our purposes, none of the initially accepted acceleration solutions were fit with orbital solutions. We thus remove from further consideration sources whose 7- or 9-parameter acceleration solutions satisfied Equations~\ref{eq:accept} or~\ref{eq:accept_7}. A total of $5.3\times 10^5$ sources were thus removed; of these, $2.4\times 10^5$ passed all cuts to be published with acceleration solutions. The properties of these sources are examined in Section~\ref{sec:accel_mock}.

Sources with \texttt{RUWE} $>1.4$ that failed the acceptance criteria for both 7- and 9-parameter acceleration solutions were next fit with orbital solutions.

\subsection{Orbital solutions}
The final astrometric model we consider is a full orbital solution. This model presumes that the 1D astrometry traces the photocenter of two bodies moving in a Keplerian orbit.  The displacement in the AL direction is given by

\begin{equation}
\begin{aligned}
 \eta = & \left[ \Delta\alpha^{*} +  \mu_{\alpha}^{*}t + B X + G Y \right] \sin \psi \\
& + \left[ \Delta\delta +  \mu_\delta t + A X + F Y \right] \cos \psi 
 + \Pi_\eta \varpi.
\end{aligned}
\label{eq:AL}
\end{equation}
Here $A$, $B$, $F$, and $G$ are the Thiele-Innes elements, which for our purposes are free parameters. The non-linear orbital parameters $P_{\rm orb}$, $e$, $\phi_{0}$ enter  via the functions
\begin{equation}
\label{eq:XY}
X = \cos E - e, \quad Y  = \sqrt{1 - e^2} \sin E, 
\end{equation}
where the eccentric anomaly $E$ is the solution to Kepler's equation,
\begin{equation}
\label{eq:kepler}
E - e \sin E = \frac{2\pi t}{P_{\rm orb}}-\phi_{0}. 
\end{equation}

The Thiele-Innes elements are transformations of the angles $\Omega$, $\omega$, $i$, and $a_0$ that can also parameterize the photocenter ellipse \citep[representing the longitude of the ascending node, argument of periastron, inclination, and photocenter semi-major axis; e.g.][]{Binnendijk1960}. As described in Appendix~\ref{sec:appendix_fitting}, Equation~\ref{eq:AL} is linear in 9 parameters and can be solved via matrix inversion once the three nonlinear parameters $P_{\rm orb}$, $e$, and $\phi_0$ are specified. Unlike with the single-star and accelerating solutions, nonlinear optimization in at least 3 parameters is required to find the best-fit parameters, and the likelihood surface is often bumpy due to period aliases. The problem is qualitatively similar to the problem of fitting a Keplerian model to a radial velocity timeseries, and the same optimizations algorithms that have been employed for that problem are also useful here. 

The fitting approach we use is described in detail in Appendix~\ref{sec:appendix_fitting}. In brief, we use adaptive simulated annealing in 3 dimensions to find the maximum-likelihood values of $P_{\rm orb}$, $e$, and $\phi_0$. We then calculate uncertainties on all 12 parameters from the Hessian matrix evaluated at the maximum-likelihood solution. Finally, we calculate the photocenter semi-major-axis $a_0$, and its uncertainty, $\sigma_{a_0}$, from the Thiele-Innes elements.
The significance of orbital solutions is quantified as the ratio of the photocenter semi-major axis to its uncertainty: 

\begin{equation}
    \label{eq:sig_a0}
    s_{\rm orb} = \frac{a_0}{\sigma_{a_0}}.
\end{equation}
Following {\it Gaia} DR3, the initial selection criteria for orbital solutions were:
\begin{equation}
\label{eq:accept_orb}
\begin{cases}
s_{\rm orb} > 5 \\
F_2 < 25 \\
\end{cases}
\end{equation}
We fit all sources within 2 kpc that satisfy UWE $> 1.4$ and were not assigned acceleration solutions with an orbital model. $5.4\times 10^5$ solutions passing the cuts in Equation~\ref{eq:accept_orb} were saved for further analysis. A majority of these sources were ultimately rejected in post-processing, as described in Section~\ref{sec:post_process}. 

Figure~\ref{fig:epoch_astrometry} shows simulated epoch astrometry of an example source. We simulate a binary with $P_{\rm orb}=902$ d, $e=0.5$, and $d=400$\,pc, located at coordinates ($\alpha,\delta$) = (153, 34) deg. We assume $M_1 = 1.0\,M_{\odot}$ and $M_2 = 0.7\,M_{\odot}$, with both components on the main sequence and a flux ratio $f=0.1$. This leads to an apparent magnitude $G=12.5$, corresponding to AL uncertainty of 0.128 mas per CCD transit, or 0.045 mas per FOV transit. The top panel shows the transit-averaged AL displacements, while the next 4 panels show residuals of these displacements with respect to the best-fit single-star model, constant and variable-acceleration models, and Keplerian orbital model. The figure illustrates a perhaps unexpected feature of the astrometric model cascade used in DR3: there are cases where a high-quality orbital solution could have been derived, but an acceleration solution is selected instead. This fact is also discussed by \citet{Halbwachs2023}; we investigate accelerations solutions further in Section~\ref{sec:accel_mock}.

\subsection{Computational cost}
We ran the full cascade of astrometric models on all 46 million unresolved binaries brighter than $G=19$ within 2 kpc. Running on a single 64-core node, this required about 80 hours of wallclock time, or a total of 5k CPU hours. The runtime is dominated by the $\sim 1\%$ of binaries for which an orbital solution is fit, since the linear models are inexpensive. Producing a similar mock catalog for DR4 data (Section~\ref{sec:dr4}) would require about 50k CPU hours, since (a) there are about twice as many data points for all sources, and (b) a larger fraction of binaries receive orbital solutions. We thus expect that it will be feasible to fit all epoch astrometry published in DR4 with a variety of models at modest, but not completely negligible, computational expense.

\begin{figure*}
    \centering
    \includegraphics[width=\textwidth]{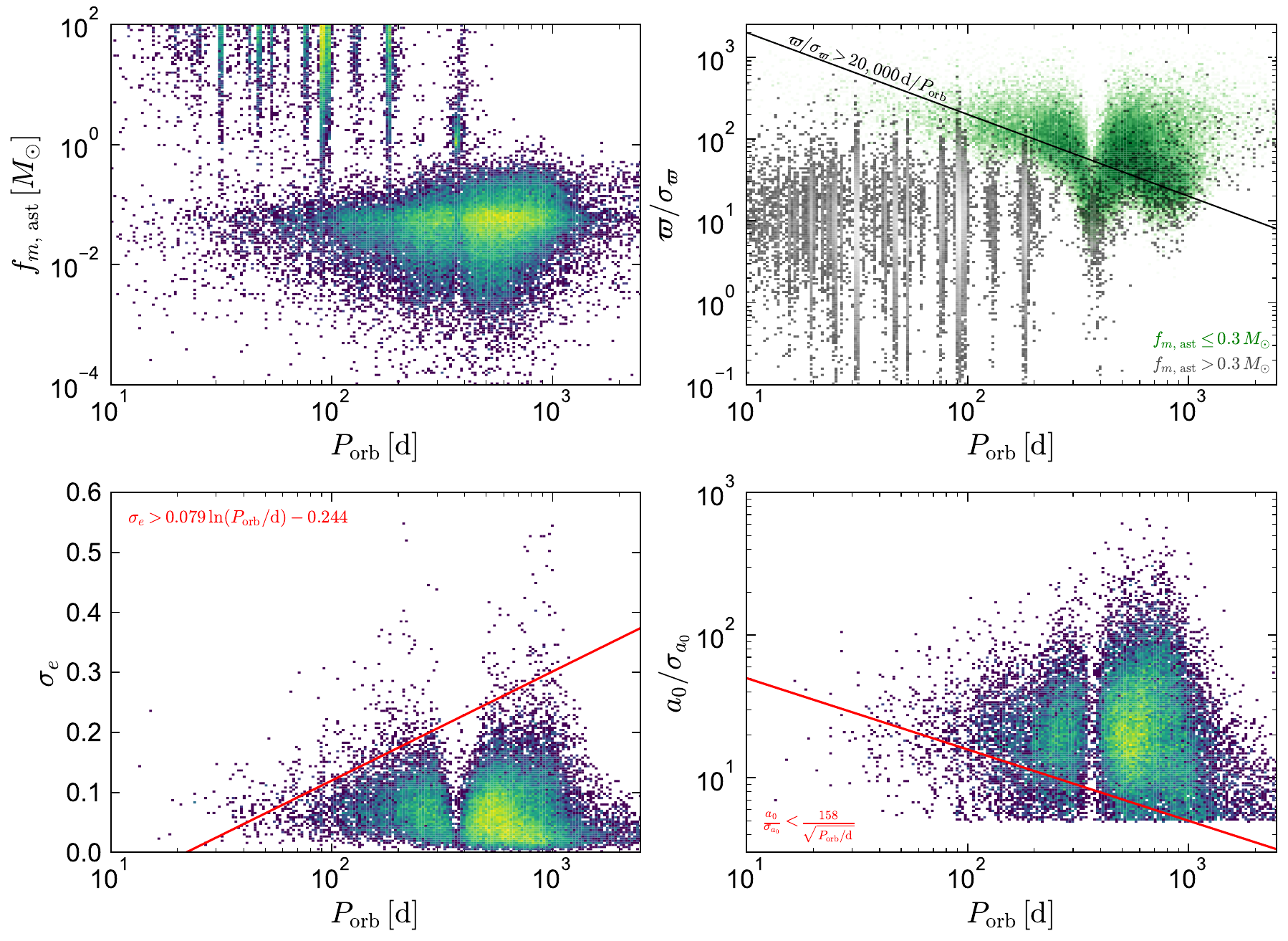}
    \caption{Selection of the astrometric solutions for our mock astrometric binary catalog. This figure in inspired by Figure 3 of \citet{Halbwachs2023}, who present the same quantities for the astrometric solutions of real sources published in {\it Gaia} DR3. Upper left panel shows all solutions with $a_0/\sigma_{a_0} > 5$ and $F_2 < 25$. The cloud of points at long $P_{\rm orb}$ and low $f_{m,\,{\rm ast}}$ consists mostly of good solutions, while the vertical stripes of solutions with high $f_{m,\,{\rm ast}}$ and short periods are all spurious. In the upper right, green and gray show the distributions of solutions with  $f_{m,\,{\rm ast}} < 0.3\,M_{\odot}$ (presumed mostly reliable) and  $f_{m,\,{\rm ast}} > 0.3\,M_{\odot}$ (presumed mostly spurious). Black line shows the empirical cut (Equation~\ref{eq:parallax_cut}) below which solutions are discarded. Bottom panels show only solutions that pass this cut. Red lines in these panels show additional cuts applied to the final solutions: only solutions below (left panel) and above (right panel) the red lines were  retained.   }
    \label{fig:halbwachs_figure}
\end{figure*}

\begin{figure*}[!ht]
    \centering
    \includegraphics[width=\textwidth]{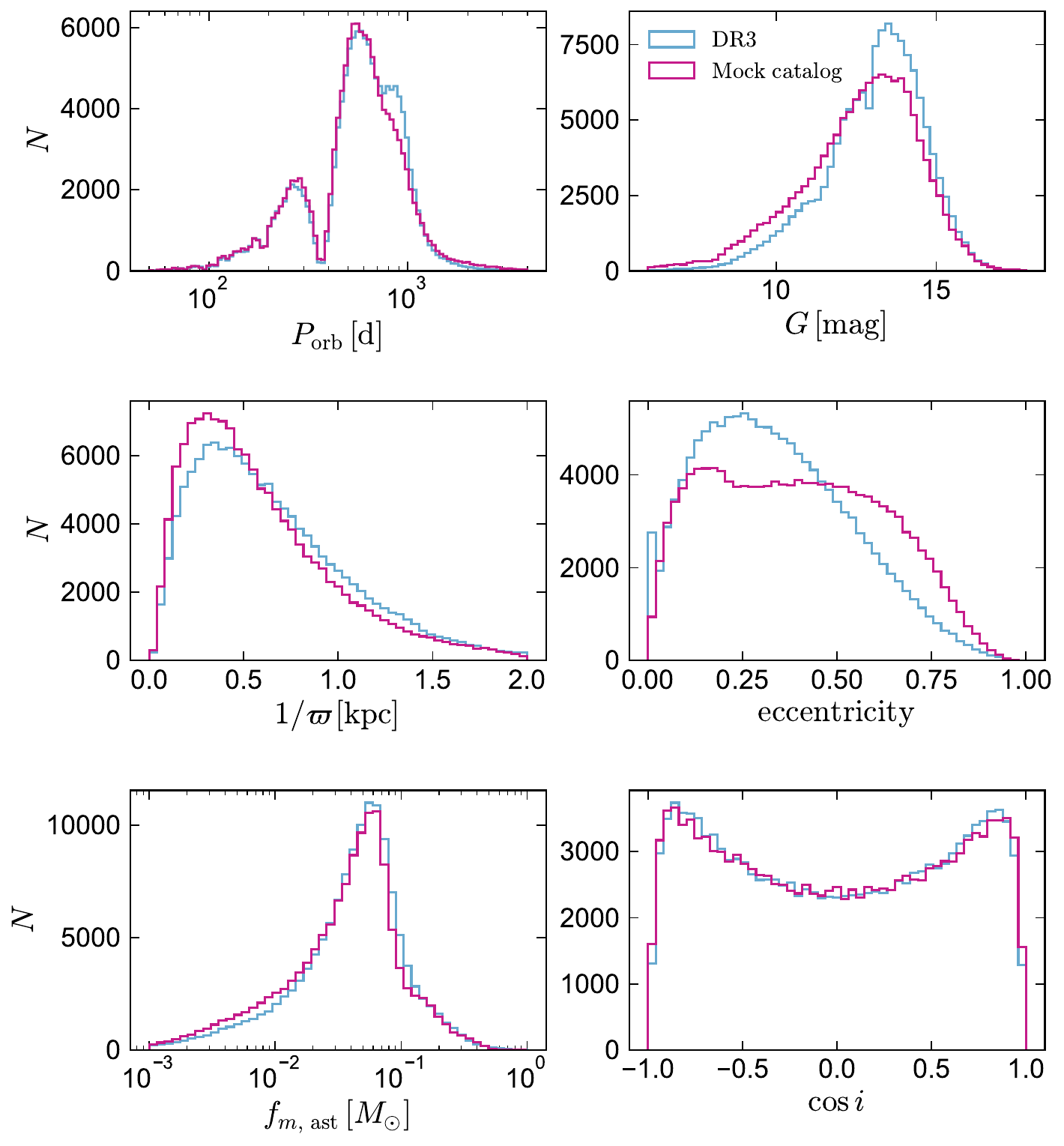}
    \caption{Comparison of the astrometric orbital solutions published in DR3 (blue) and the mock catalog after cuts (violet). Each panel shows the distribution of one parameter, marginalized over all other parameters. The mock catalog's distributions of orbital period, apparent magnitude, distance, inclination, and astrometric mass function are all in reasonably good agreement with the DR3 data. The mock catalog predicts more eccentric orbits than are observed, suggesting the input eccentricity distribution has too many high-eccentricity orbits. }
    \label{fig:catalog_comparison}
\end{figure*}

\subsection{Post-processing}
\label{sec:post_process}
Figure~\ref{fig:halbwachs_figure} shows properties of the provisionally-accepted orbital solutions satisfying Equation~\ref{eq:accept_orb}. The figure is inspired by Figure 3 of \citet{Halbwachs2023} and shows the same quantities for our simulated solutions as that figure shows for actual solutions calculated with DR3 data. Only 20\% of the data are shown in each panel, and the color scale is logarithmic.

For each solution, we calculate the astrometric mass function, 
\begin{equation}
    \label{eq:fm_ast}
    f_{m,\,{\rm ast}}=\left(\frac{a_{0}}{\varpi}\right)^{3}\left(\frac{P_{{\rm orb}}}{1\,{\rm yr}}\right)^{-2}.
\end{equation}
This quantity is related to the companion's mass and is also useful for distinguishing between real and spurious solutions. For a binary containing a dark companion, $f_{m,\,{\rm ast}}$ represents the mass in solar units required to explain the observed orbit if the luminous star is a massless test particle. Because the photocenter orbit in a luminous binary is significantly smaller than the true semi-major axis -- and because the luminous stars are {\it not} massless -- $f_{m,\,{\rm ast}}$ is usually significantly lower than the true companion mass \citep[e.g.][]{Shahaf2019}. The largest mass function expected for binaries containing two main sequence stars is about $0.1\,M_\odot$. Binaries in which the primary is a giant or the secondary is itself a close binary can reach $f_{m,\,{\rm ast}} \sim 0.3\,M_{\odot}$. Systems containing WDs can approach $f_{m,\,{\rm ast}} \sim 1\,M_{\odot}$ in the most extreme cases, but this requires a low-mass primary and a massive WD, so most WD binaries are also confined to $f_{m,\,{\rm ast}} \lesssim 0.3\,M_\odot$. In summary, a large majority of astrophysical binaries are expected to have $f_{m,\,{\rm ast}} \lesssim 0.3\,M_\odot$. Orbits with significantly larger $f_{m,\,{\rm ast}}$ are thus either spurious or contain black holes and neutron stars.

The upper left panel of Figure~\ref{fig:halbwachs_figure} shows the orbital periods and mass functions of binaries initially accepted with orbital solutions. A cloud of solutions with $f_{m,\,{\rm ast}}\lesssim 0.3\,M_\odot$ and $100\lesssim P_{{\rm orb}}/{\rm d}\lesssim1000$ d is evident, consisting mostly of binaries with well-measured orbital solutions. In addition, there is a secondary population of solutions with $P_{\rm orb}$ mostly falling in several narrow vertical stripes and mass functions extending to much larger values, including well beyond the limits of the plot. As we explore further in Section~\ref{sec:spurious}, these are almost all wide binaries -- with typical true periods of 100-1000 years -- and spurious apparent astrometric motion on periods related to the observing cadence. A wide range of periods are present: the most common values are 91, 31.5, and 182 days, but some spurious solutions exist at almost all periods.

\begin{figure*}
    \centering
    \includegraphics[width=\textwidth]{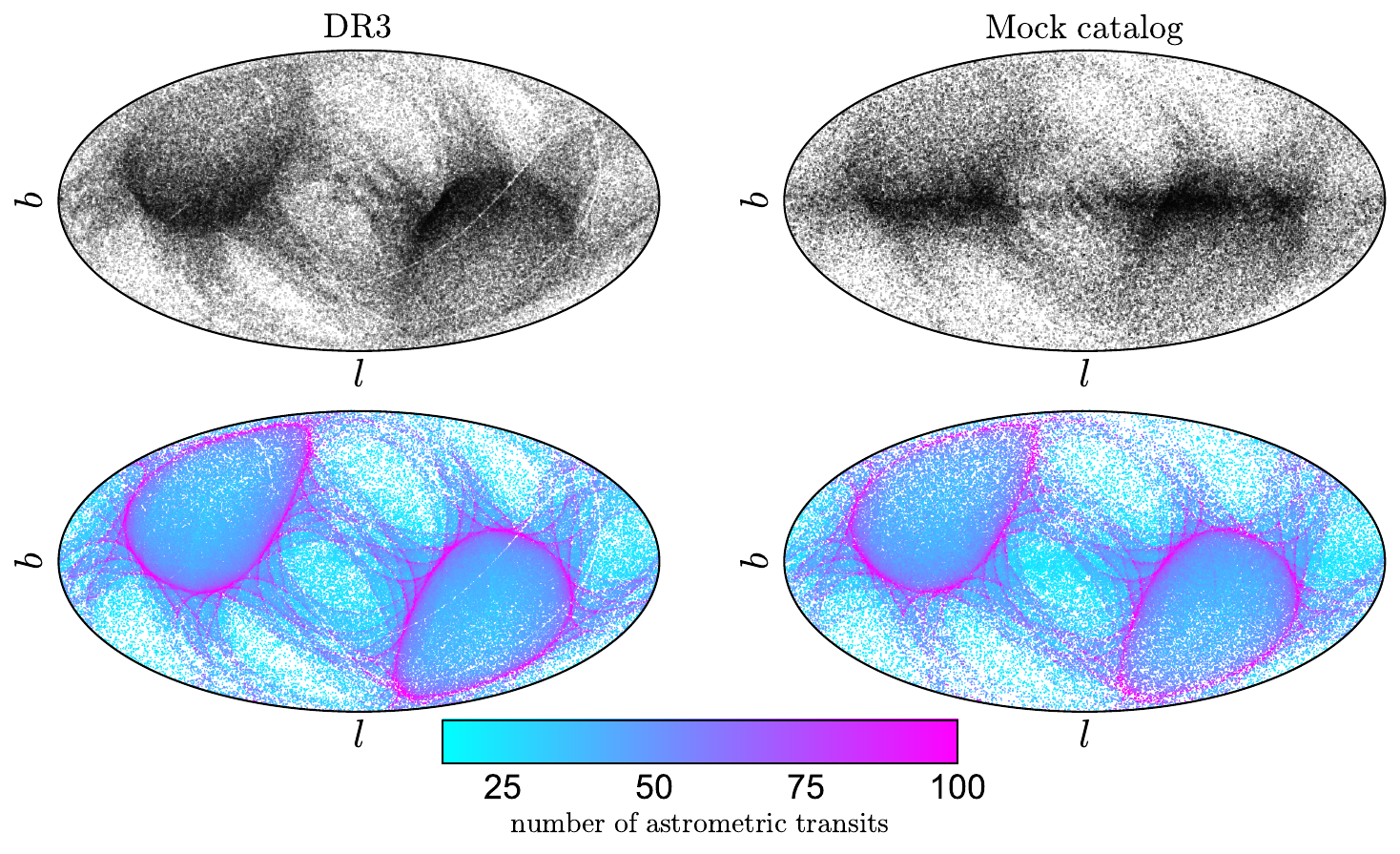}
    \caption{Galactic sky distribution of binaries in the DR3 astrometric binary catalog (left) and our mock catalog (right). The Galactic center is in the middle of the projection. In the bottom panels, sources are colored by the number of astrometric transits during the DR3 observing window. Imprints of the {\it Gaia} scanning law are clearly visible in both catalogs but are sharper in the the observed catalog. A few narrow ribbons in which there are no sources in the DR3 catalog are not present in the mock catalog. We believe these are a result of poor quality astrometry obtained near decontamination events, which was not filtered out in producing the DR3 catalog. }
    \label{fig:catalog_comparison_sky}
\end{figure*}

\citet{Holl2023b} discuss the origin of these periods in detail. The goodness-of-fit (Equation~\ref{eq:F2}) of the spurious solutions in our simulations is often unremarkable compared to the good solutions, and there is no simple way to reliably distinguish all of them from solutions of genuine binaries that happen to have the same periods. However, \citet{Halbwachs2023} found that most of the spurious solutions can be removed from the sample with a period-dependent cut on the fractional parallax uncertainty, $\varpi/\sigma_{\varpi}$. The cut they used is illustrated in the upper right panel of Figure~\ref{fig:halbwachs_figure}:
\begin{equation}
    \label{eq:parallax_cut}
    \frac{\varpi}{\sigma_{\varpi}} > \frac{20,000\,\rm d}{P_{\rm orb}}.
\end{equation}
Following \citet{Halbwachs2023}, we color sources with $f_{m,\,{\rm ast}} < 0.3\,M_{\odot}$ (mostly good solutions) in green in Figure~\ref{fig:halbwachs_figure}, and sources with $f_{m,\,{\rm ast}} > 0.3\,M_{\odot}$ (mostly spurious solutions) in gray. A majority of the spurious solutions fall below the black line and are thus filtered out by Equation~\ref{eq:parallax_cut}, as was also the case with the real DR3 data.

Filtering of the spurious solutions comes at the expense of a significant number of good orbital solutions for real binaries. In total, 46\% of all the green points in the right panel of Figure~\ref{fig:halbwachs_figure} fall below the dashed line and are excluded.  As expected, binaries with periods near one year have large parallax uncertainties -- because their orbital motion is degenerate with parallactic motion -- and are almost all eliminated by Equation~\ref{eq:parallax_cut}. This is desirable, since the left panel of Figure~\ref{fig:halbwachs_figure} shows that most such binaries have vastly overestimated mass functions. Binaries with periods close to 1 year are more likely to be detected if they are nearby or if they have eccentric orbits, because eccentricity breaks the degeneracy between parallax and orbital motion. 

The bottom panels of Figure~\ref{fig:halbwachs_figure} show the solutions that remain after removal of sources not satisfying Equation~\ref{eq:parallax_cut}. The majority of the spurious solutions at short periods have been removed, but some vertical stripes associated with scan angle-dependent periods are still apparent. The panels show the eccentricity uncertainty (lower left) and the photocenter semi-major axis divided by uncertainty (i.e., the significance; lower right). Red lines show two additonal cuts imposed by \citet{Halbwachs2023} in post-processing: 

\begin{align}
\label{eq:filters}
\sigma_{e} &<0.079\ln\left(P_{{\rm orb}}/{\rm d}\right)-0.244 \\ 
\label{eq:filter2}
\frac{a_{0}}{a}&>\frac{158}{\sqrt{P_{{\rm orb}}/{\rm d}}}.
\end{align}

These filters were chosen empirically because most of the manifestly spurious solutions fall above and below them, respectively. We retain only the simulated solutions that pass both cuts. After Equation~\ref{eq:parallax_cut} has been applied, only a few percent of the surviving solutions are removed by these cuts. However, Figure~\ref{fig:halbwachs_figure} shows that the filters are successful in removing the visible band of spurious solutions at $P_{\rm orb} \approx 91$ d that survived Equation~\ref{eq:parallax_cut}.

\section{The final mock binary catalog}
\label{sec:mock_catalog}
The cuts described above leave us with a catalog of 137,000 astrometric orbital solutions, which can be compared to the 168,000 solutions published in {\it Gaia} DR3.\footnote{We note that 33467 of the astrometric orbital solutions published in DR3 are \texttt{AstroSpectroSB1}, meaning that they were computed from a combination of astrometric and RV data. We have not attempted to forward-model the {\it Gaia} RV data. However, having an accepted \texttt{Orbital} solution was a necessary prerequisite for being processed with an  \texttt{AstroSpectroSB1} solution in DR3, so we expect our modeling to still be largely applicable to the sources with these solutions.} The mock catalog is 18\% smaller. This modest mismatch can likely be attributed to uncertainties in the underlying stellar and binary population. For example, we show below that the modeled population likely has more eccentric orbits than the true population and that {\it Gaia} is less sensitive to high-eccentricity orbits. 


We compare the properties of the final simulated and observed catalogs in the following sections. To facilitate clear comparison of the distributions of binary properties between the observed and simulated catalogs, we discard a random 19\% of the observed catalog in these comparisons, such that the simulated comparison sample has the same number of binaries as the observed catalog. 

\subsection{Comparison to the DR3 catalog}

Figure~\ref{fig:catalog_comparison} compares the properties of the observed and simulated catalogues of orbital solutions. From upper left to lower right, panels show distributions of orbital period, apparent $G-$band magnitude, distance, eccentricity, astrometric mass function (Equation~\ref{eq:fm_ast}) and inclination. Most of these parameters have similar distributions in the observed and mock catalogs. 

\begin{figure*}
    \centering
    \includegraphics[width=\textwidth]{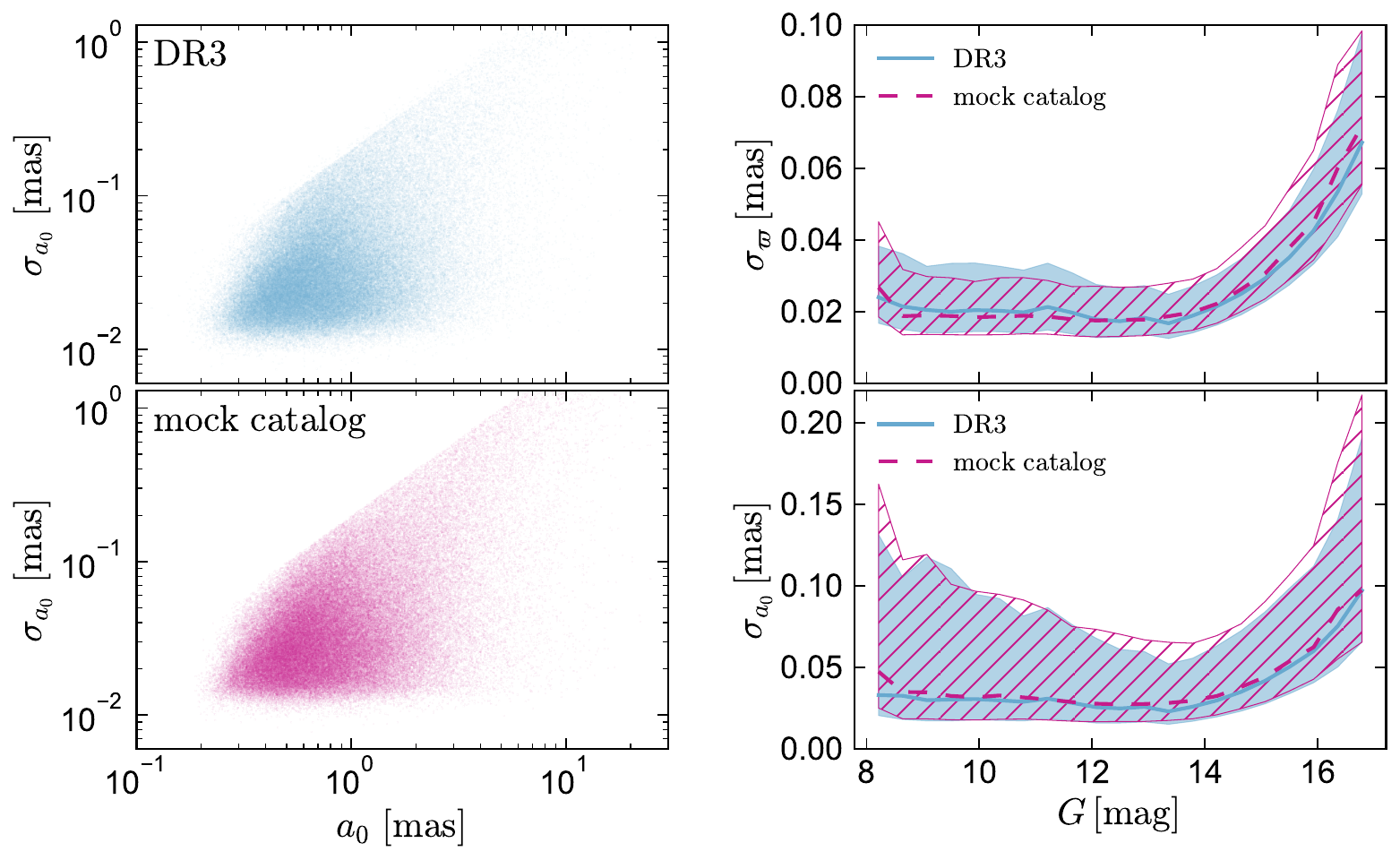}
    \caption{Left panels: Photocenter semi-major axis and uncertainty of the {\it Gaia} DR3 astrometric orbital solutions (top) and solutions in the our mock catalog (bottom). In both sets of solutions, $\sigma_{a_0}$ is correlated with $a_0$, but this is primarily a selection effect owing to the cut on $s=a_0/\sigma_{a_0}$. Right panels: uncertainty in parallax (top) and photocenter semi-major axis (bottom) as a function of $G-$band apparent magnitude. We show the median and middle 68\% uncertainty range for both the real and synthetic catalogs. The agreement is generally quite good.  }
    \label{fig:a0}
\end{figure*}


Both the observed and simulated catalogs have period distributions that peak at $\sim 550$\,d, with a sharp dip at 1 yr and a smaller dip at 0.5 yr. This structure is almost entirely a result of the selection function, since the assumed underlying selection function is close to log-flat (but rising gradually toward longer periods; see Section~\ref{sec:context}) over the relevant period range. The apparent magnitude distributions of the two catalogs are quite similar, though a discontinuity at $G=13$ is more pronounced in the observed sample. The simulated sample peaks at a slightly closer distance and has slightly more low-$f_{m,\,\rm ast}$ solutions. The bias against edge-on orbits is reproduced well in the simulated catalog.

The most significant difference between the two catalogs is in their eccentricity distributions, where the mock catalog has significantly higher typical eccentricities. It seems most plausible that this simply reflects an imperfect input binary population assumed in our simulations. For solar-type primaries, which dominate the sample, the eccentricity distribution is taken from \citet{Moe2017} and assumes $p(e) \propto e^{0.2}$ at periods of 100-1000 d.  That is, the input eccentricity distribution is more eccentric than uniform. This distribution was inferred by \citet{Moe2017} by fitting the eccentricities of 15 solar-type binaries with $P_{\rm orb}=100-1000$ d from the \citet{Raghavan2010} sample. However, only three of these binaries have $e>0.6$, and we consider it likely that the true eccentricity distribution has less support at high eccentricities than $p(e) \propto e^{0.2}$. Since {\it Gaia} is less sensitive to high-eccentricity orbits, this may also explain why the simulated catalog is smaller than the observed one. Refining the inferred population statistics of binaries in this period range using the {\it Gaia} sample is a promising avenue for future work.

\subsubsection{Sky distribution}
Figure~\ref{fig:catalog_comparison_sky} compares the sky distribution of binaries in the observed and mock catalog. Like the observed catalog, the mock catalog has an overdensity of binaries in the Galactic plane, but an underdensity toward the Galactic center. This is mainly a consequence of the {\it Gaia} scanning law: the bottom panels of the figure show the number of individual FOV transits included in the astrometric solution for each source during the DR3 observing window. The number of FOV transits per source in the mock catalog matches the observed catalog with high fidelity, since the \texttt{GOST}-predicted scanning law is used in simulating our mock observations. Curiously, imprints of the scanning law in the observed catalog are somewhat more pronounced than in the mock catalog, perhaps suggesting that the quality of the {\it Gaia} observations and/or treatment of systematics varies with position on the sky. 

Several narrow ribbons on the sky with essentially no good solutions are also evident in the observed catalog. No such features are present in the mock catalog, indicating that their cause is not included in our modeling. We found that these features  all correspond to regions of the sky observed just before or after several ``decontamination'' events in which the {\it Gaia} mirrors and CCDs were heated to remove accumulated ice \citep[e.g.][]{Lindegren2021}. Astrometric measurements obtained close to decontamination events are expected to have poor quality and were discarded when calculating the single-star astrometric solutions reported in the \texttt{gaia\_source} table. We suspect that these measurements were unintentionally included when calculating orbital solutions, and that the presence of a single spurious astrometric measurement prevented the calculation of a good solution in regions observed close to decontamination events.

\subsubsection{Astrometric uncertainties}
Figure~\ref{fig:a0} compares the astrometric uncertainties of the observed and mock catalogs. Recall that these quantities are calculated from the Hessian matrix; i.e., from the curvature of the likelihood function in the vicinity of the best-fit solution. They thus depend mainly on the assumed epoch-level astrometric uncertainties and number of scans per source. The left panels show the photocenter semi-major axis, $a_0$, and its uncertainty, $\sigma_{a_0}$. As been pointed out in other work \citep[e.g.][]{Shahaf2023, El-Badry2023}, these two quantities are correlated in the observed catalog. Such a correlation may not be naively expected, since $\sigma_{a_0}$ for a given set of orbital parameters and scan cadence is expected to depend only on the epoch-level astrometric uncertainties in the limit where a whole orbit is observed. The same correlation is present in the mock catalog. The correlation arises because of the cut of $s_{\rm orb}>\max\left(5,\frac{158}{\sqrt{P_{{\rm orb}}/{\rm d}}}\right)$ employed in constructing the catalog, which removes sources in the upper left of the $\sigma_{a_0}$ vs. $a_0$ plane.

\subsubsection{RUWE}
\label{sec:ruwe}

\begin{figure}
    \centering
    \includegraphics[width=\columnwidth]{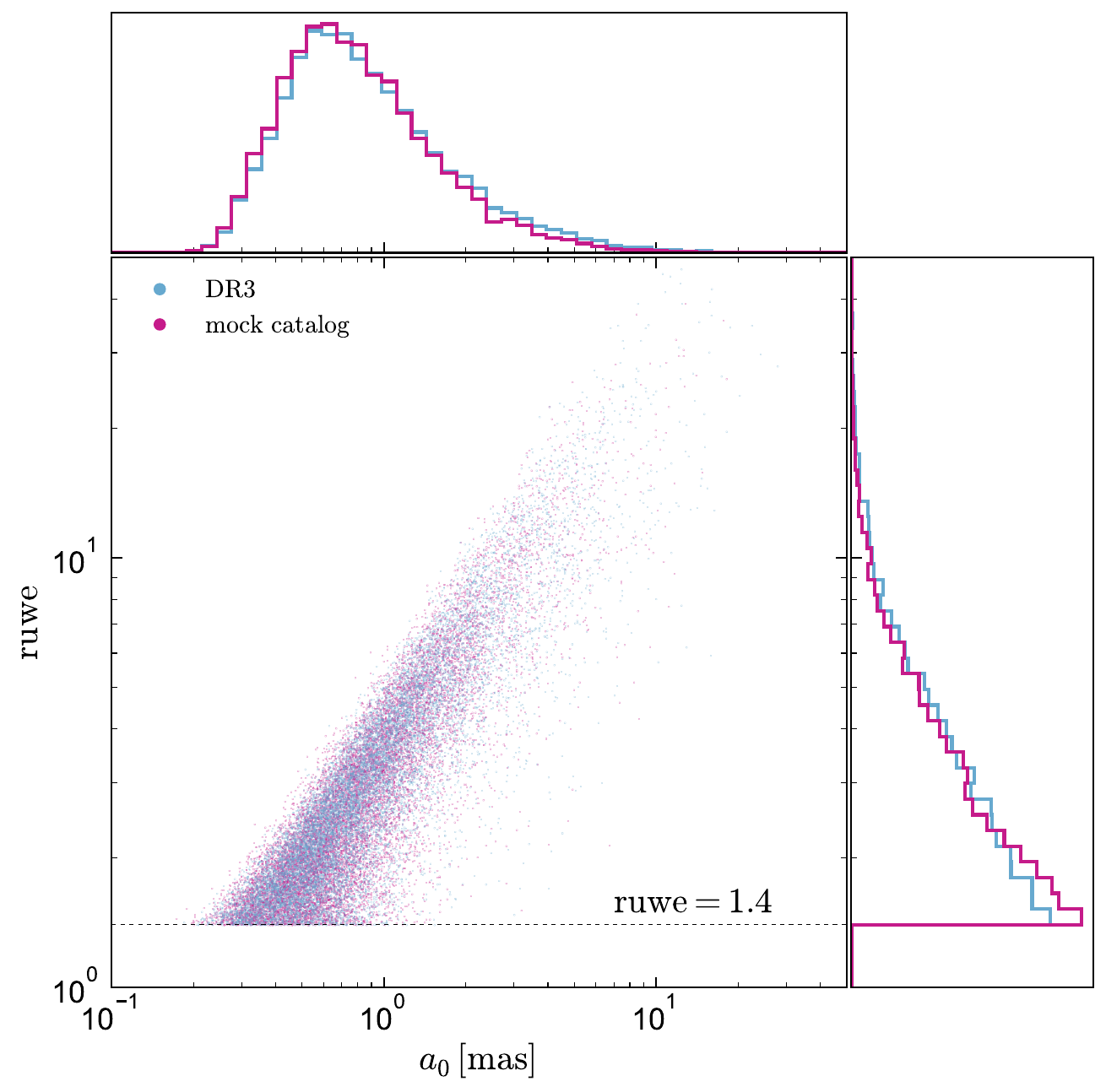}
    \caption{Photocenter semi-major axis, $a_0$, and \texttt{RUWE} (Equation~\ref{eq:uwe}) for the DR3 astrometric binary catalog (blue) and the mock catalog (violet). A random 10\% of solutions are plotted. $a_0$ is calculated from the 12-parameter orbital solutions, while \texttt{RUWE} is calculated from the 5-parameter single-star solution. The scatter in \texttt{RUWE} at fixed $a_0$ is driven primarily by scatter in eccentricity and inclination. The mock catalog's good agreement with observations suggests that the adopted noise model is reasonably accurate. }
    \label{fig:ruwe}
\end{figure}

Figure~\ref{fig:ruwe} compares the observed and simulated \texttt{RUWE} values and photocenter semi-major axes. As shown by \citet{Casey}, these parameters are strongly correlated, with most of the scatter due to the range of eccentricities and inclinations in the catalog. No solutions with \texttt{RUWE} $<1.4$ are shown, because these are not fit with orbital solutions. The model reproduce the relation and its scatter well, suggesting that our treatment of the 5-parameter solutions and noise model is a good approximation of reality.

\subsection{Spurious solutions}
\label{sec:spurious}

\begin{figure*}
    \centering
    \includegraphics[width=\textwidth]{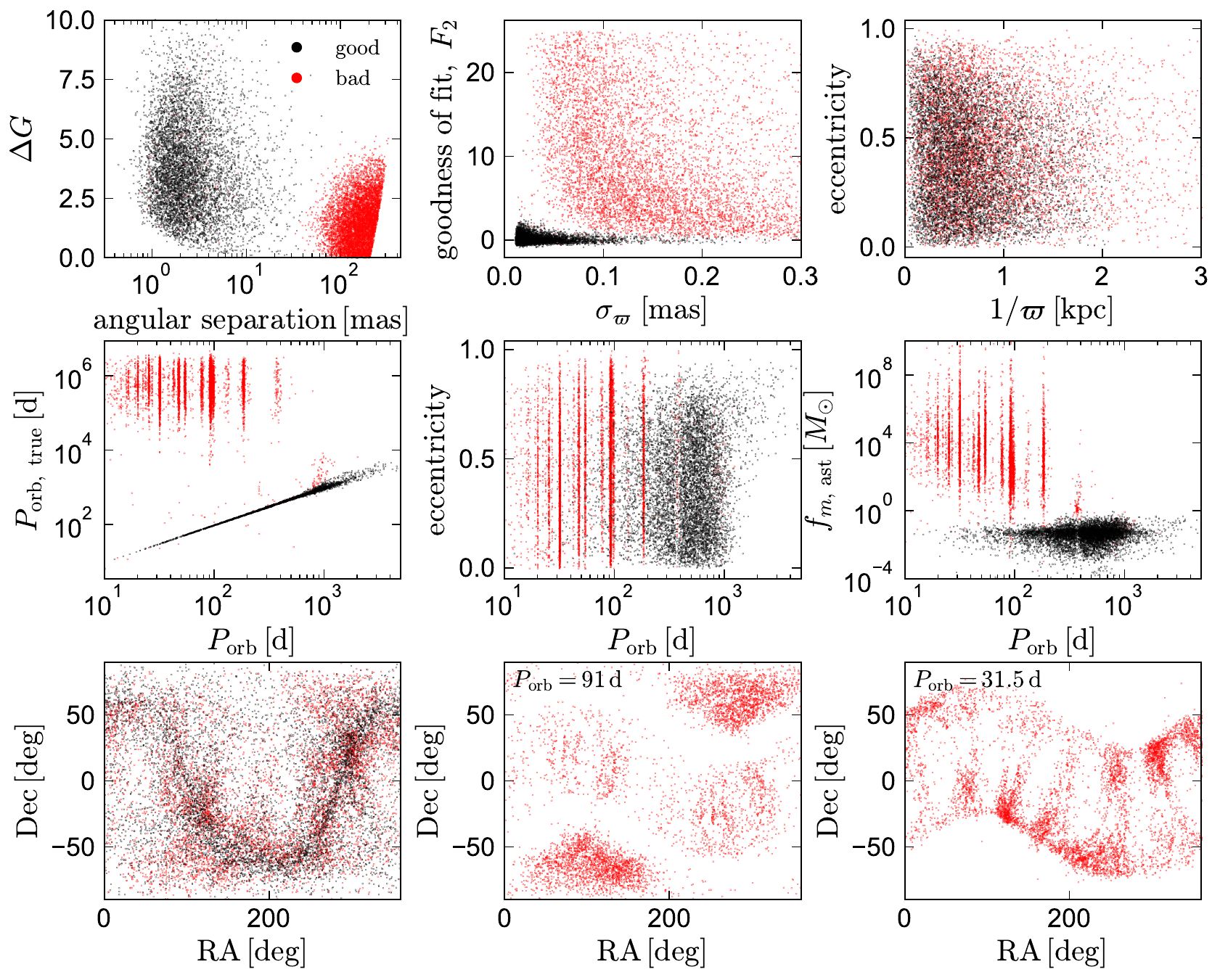}
    \caption{Comparison of good (black) and spurious (red) astrometric solutions, before post-processing. We show a random subset of $10^4$ solutions in each category. We define spurious solutions as those for which the inferred period differs from the true period by at least $5\sigma$. Upper left shows true angular separation and magnitude difference between the two stars. A large majority of the bad solutions are marginally resolved pairs with separations of order 100 mas. The inferred eccentricities, parallaxes, parallax uncertainties, and goodness of fit values of these solutions largely overlap with the good solutions. Almost all of the spurious solutions have inferred periods of a year or less -- mostly related to the scanning law -- while 80\% of the true solutions have longer inferred periods. Most, but not all, of the bad solutions have unphysically large mass functions. Bottom panels shows the sky distributions. Spurious solutions are found across most of the sky, but spurious solutions of different periods are found in different locations. }
    \label{fig:spurious}
\end{figure*}

Figure~\ref{fig:spurious} investigates the spurious solutions found in the initial mock catalog, most of which are ultimately removed by the post-processing cuts. We define spurious solutions as those whose inferred periods are inconsistent with their true periods by at least $5\sigma$, and good solutions as all others that satisfy $s_{\rm orb} > 5$ and $F_2 < 25$.  We plot a random subset of $10^4$ solutions in each class. These are the same solutions shown in Figure~\ref{fig:halbwachs_figure} before the post-processing cuts (Equations~\ref{eq:parallax_cut}-\ref{eq:filter2}) are applied.

Most of the spurious solutions have true periods of decades to millenia and angular separations of 50-200 mas. These binaries experience little actual orbital motion during the DR3 observing baseline, but their marginally resolved separations combined with the one-dimensional nature of {\it Gaia} astrometry gives rise to spurious scan-angle dependent signals (Appendix~\ref{sec:appendix_resolved}) with a variety of periods related to the scanning law  \citep[e.g.][]{Holl2023b}.

In addition to the large population of spurious solutions related to marginally resolved sources, there are also some spurious solutions with shorter true periods. These include (1) a small population of sources with inferred $P_{\rm orb}\approx 1000$ d and true $P_{\rm orb} =10^{3-4}\,{\rm d}$ -- for which the uncertainties inferred from linearized fitting are likely unreliable -- and (2) a small number of sources with shorter periods for which fitting simply yielded the wrong period. It is possible that the number of spurious solutions with true periods $< 1000$ d could be reduced through improvements to the optimization algorithm used in orbit fitting (Appendix~\ref{sec:appendix_fitting}), but since these solutions are few in number and are almost all removed in post-processing, we do not concern ourselves with them further here. 

Given their short periods, most of the spurious solutions have very large mass functions, and could be eliminated with a cut of $f_{m,\,{\rm ast}} < 1\,M_{\odot}$. This approach is, however, not desirable, because the small number of good solutions with high mass functions are of significant astrophysical interest. Similarly, since spurious solutions are primarily clustered around particular periods, most of them could be eliminated by filtering out solutions with these periods. A related approach, which is advocated by \citet{Holl2023b}, would be to filter out solutions for which the AL displacements can be well-fit by a sinusoidal function of the scan angle. However, this would come at the expense of some true orbits that happen to have periods close to the problematic periods associated with the scanning law. 

\begin{figure}
    \centering
    \includegraphics[width=\columnwidth]{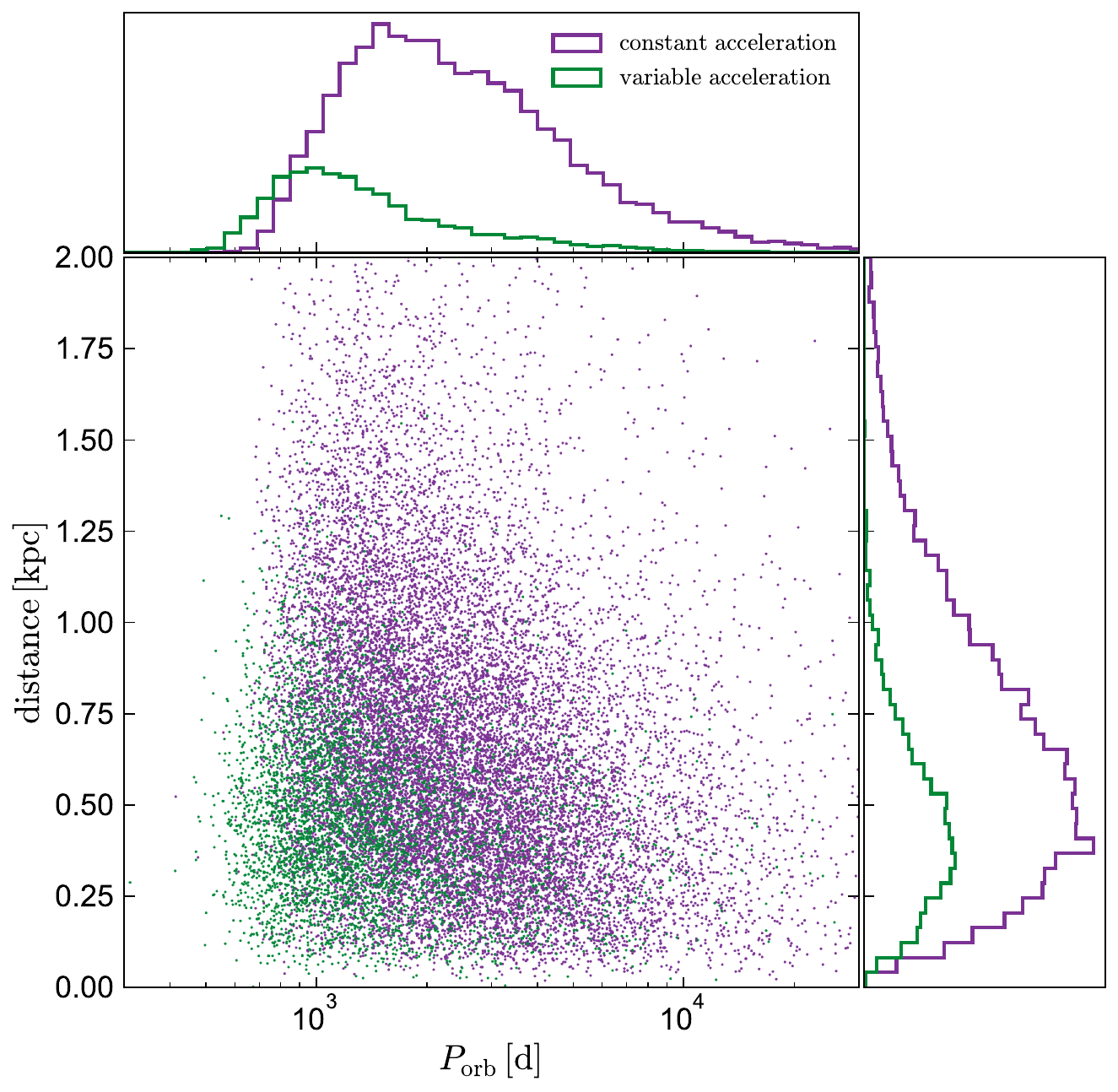}
    \caption{Orbital periods and distances of the accelerating solutions in the mock catalog. We show 10\% of all solutions. The period distributions of sources with variable-acceleration and constant-acceleration solutions peak at 1000 and 1500 d, respectively, but both extend to periods of $>10^4$ d. The fraction of solutions with true periods below 1000 d is 32\% and 6\%, respectively. }
    \label{fig:accel}
\end{figure}

The bottom panels of Figure~\ref{fig:spurious} shows the distribution of good and bad solutions on the plane of the sky. Although spurious solutions are found across most of the sky, their distribution is different from that of good solutions, and spurious solutions of different periods are concentrated at different positions on the sky \citep[see also][their Figures C.6 and C.7]{Holl2023b}.

Figure~\ref{fig:spurious} suggests that -- since true and spurious solutions have different distributions of many parameters, but do not completely separate in any parameter -- a classifier could be devised to distinguish between good and bad solutions with higher purity and completeness than the cuts used in DR3. Indeed, \citet{Rybizki2022} designed such a classifier for single-star solutions, which achieved significantly better performance than cuts on any simple combination of astrometric quality flags. It would be straightforward to train a similar classifier for binary solutions, and astrometry from a mock catalog may form a useful component of the training set. 

\subsubsection{Accelerating solutions}
\label{sec:accel_mock}
As discussed in Section~\ref{sec:cascade}, all sources with 9- or 7-parameter acceleration solutions satisfying Equations~\ref{eq:accept} and~\ref{eq:accept_7} were removed from processing and thus were ineligible to receive an orbital solution. However, additional cuts were imposed on the solutions actually published in DR3. Only solutions with significance $s > 20$ were published, even though solutions with $s > 12$  were removed from processing and not fit with an orbital model.  In addition, the $F_2$ threshold for published 7-parameter solutions was lowered from 25 to 22.

Figure~\ref{fig:accel} shows properties of sources that would have been published with acceleration solutions in the mock catalog. 
As one might expect, a majority of the sources with acceleration solutions have orbital periods longer than the DR3 observing baseline, and sources with 7-parameter solutions have longer periods on average than those with 9-parameter solutions. However, the period distributions are quite broad, and both sets of solutions have a short-period tail extending to $P_{\rm orb} < 1000$ d, with the shortest-period 9-parameter solutions having periods of order 500 d. The presence of this short-period tail will complicate any attempts to infer companion masses for sources with acceleration solutions. 
A majority of these short-period solutions would have passed the post-processing cuts with full orbital solutions. In future data releases, it will be possible to fit their epoch astrometry with orbital solutions. 

\subsection{The NSS astrometric sample in context}
\label{sec:context}

\begin{figure*}
    \centering
    \includegraphics[width=\textwidth]{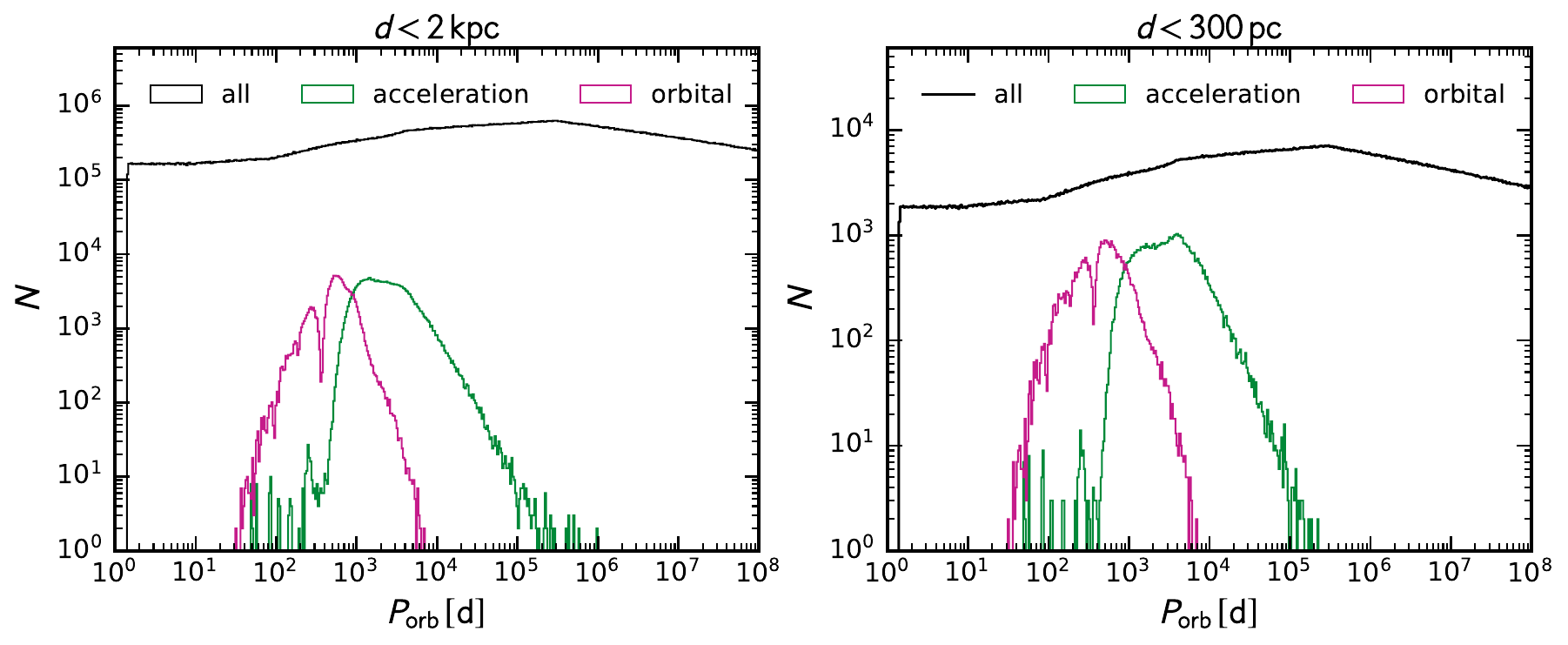}
    \caption{True orbital periods of sources accepted with orbital solutions (magenta) and acceleration solutions (green). Black histogram shows the modeled underlying period distribution of all binaries. Left panel shows sources within 2 kpc, while right panel shows those within 300 pc.  Most of the accepted solutions have true periods between $10^2$ and $10^4$ d. Even in this period range, the DR3 catalogs represent only $\approx 1\%$ of all the binaries within 2 kpc. Within 300 pc, the completeness peaks at $\sim 25\%$ at $P_{\rm orb} \approx 600$\,d. } 
    \label{fig:period_hist}
\end{figure*}

Figure~\ref{fig:period_hist} compares the periods  of sources that receive orbital and acceleration solutions to the underlying period distribution of all binaries in the model within 2 kpc (left) and 300 pc (right). Only a small fraction of all simulated binaries receive astrometric binary solutions of any kind. At $P_{\rm orb} \approx 550$\,d, where the period distribution of orbital solutions peaks, the completeness is about 2\% within 2 kpc, and 25\% within 300 pc. At $P_{\rm orb} = 100$\,d, it is only 0.03\% within 2 kpc, and $\approx 5\%$ within 300 pc. Sensitivity to shorter- and longer-period binaries in DR3 is higher for spectroscopic (SB1/SB2) orbits and resolved binaries, respectively, but the completeness of these samples remains to be characterized. 

\subsection{Predictions for DR4 and DR5}
\label{sec:dr4}
To make predictions for the {\it Gaia} DR4 binary sample, we include observations predicted by \texttt{GOST} before 20 January, 2020, corresponding to 1977 days of observations. For DR5, we include all predicted observations until January 2025. We assume that the epoch-level astrometric uncertainties and unmodeled additional noise  will be the same as in DR3 -- likely making these predictions conservative, since treatment of systematics is expected to improve in future releases. We also employ the same selection criteria and post-processing cuts (Equations~\ref{eq:parallax_cut}-\ref{eq:filter2}), though these may be relaxed in future releases.

\begin{figure*}
    \centering
    \includegraphics[width=\textwidth]{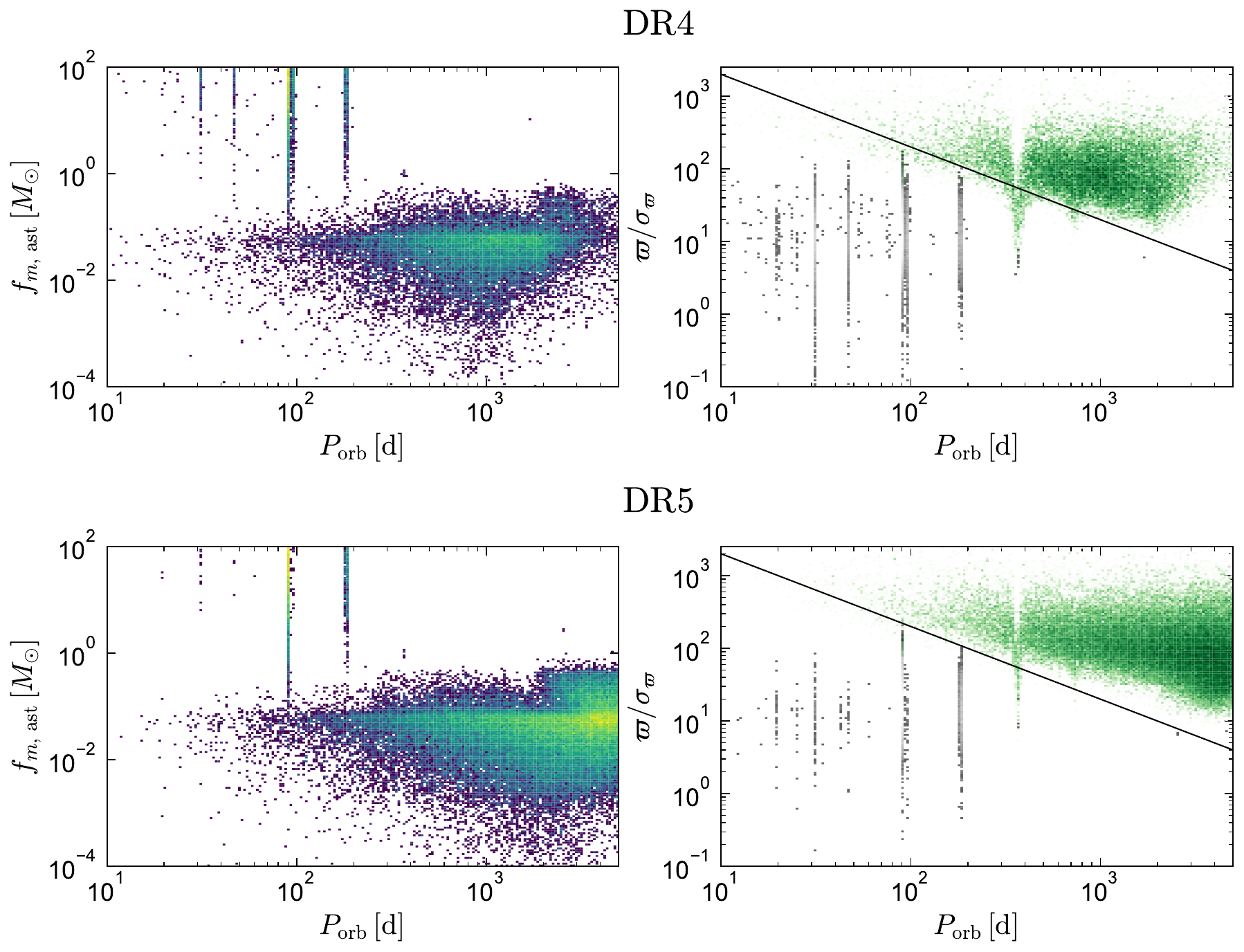}
    \caption{Analogous to the top panels of Figure~\ref{fig:halbwachs_figure}, but for {\it Gaia} DR4 (top) and DR5 (bottom). Note that the x-axis extends to longer periods than in Figure~\ref{fig:halbwachs_figure}. We show 3\% of all solutions. The number of spurious periods associated with the scanning law is predicted to decrease compared to DR3, while the parallaxes uncertainties will decrease and sensitivity to long orbital periods will increase. Most solutions satisfying $a_0/\sigma_{a_0}$ in DR5 also fall above   the $\varpi/\sigma_\varpi$ cut employed in DR3 (Equation~\ref{eq:parallax_cut}). }
    \label{fig:halbwachs_dr4}
\end{figure*}

Figure~\ref{fig:halbwachs_dr4} shows results of simulations including observations with the DR4 (top) and DR5 (bottom) time baselines. Comparing to Figure~\ref{fig:halbwachs_figure}, it is clear that (a) most good solutions will be available at longer orbital periods in future releases, and (b) there will be significantly fewer spurious solutions. The smaller number of spurious periods likely owes in part to a reversal in {\it Gaia's} precession during 2019-2020 \citep[including the last 6 months of data included in DR4;][]{Lindegren2021}, which eliminates some spurious periods. The ``hole'' near $P_{\rm orb}\approx 1$ yr is also predicted to become progressively narrower in DR4 and DR5. 

In the lower left panel of Figure~\ref{fig:halbwachs_dr4}, a ridge of higher-mass function binaries can be seen at $P_{\rm orb}\gtrsim 2000$ d. This is made up primarily of binaries containing a WD, which produce large photocenter orbits and are predicted to be common at long periods, where they can escape a common envelope event during the late-stage evolution of the WD progenitor. To avoid flagging these orbits as spurious solutions, we increase the threshold above which solutions in the right panels of Figure~\ref{fig:halbwachs_dr4} are colored gray to $f_{m,\,{\rm ast}}= 1\,M_{\odot}$, compared to $0.3\,M_{\odot}$ in Figure~\ref{fig:halbwachs_figure}. 

\begin{figure}
    \centering
    \includegraphics[width=\columnwidth]{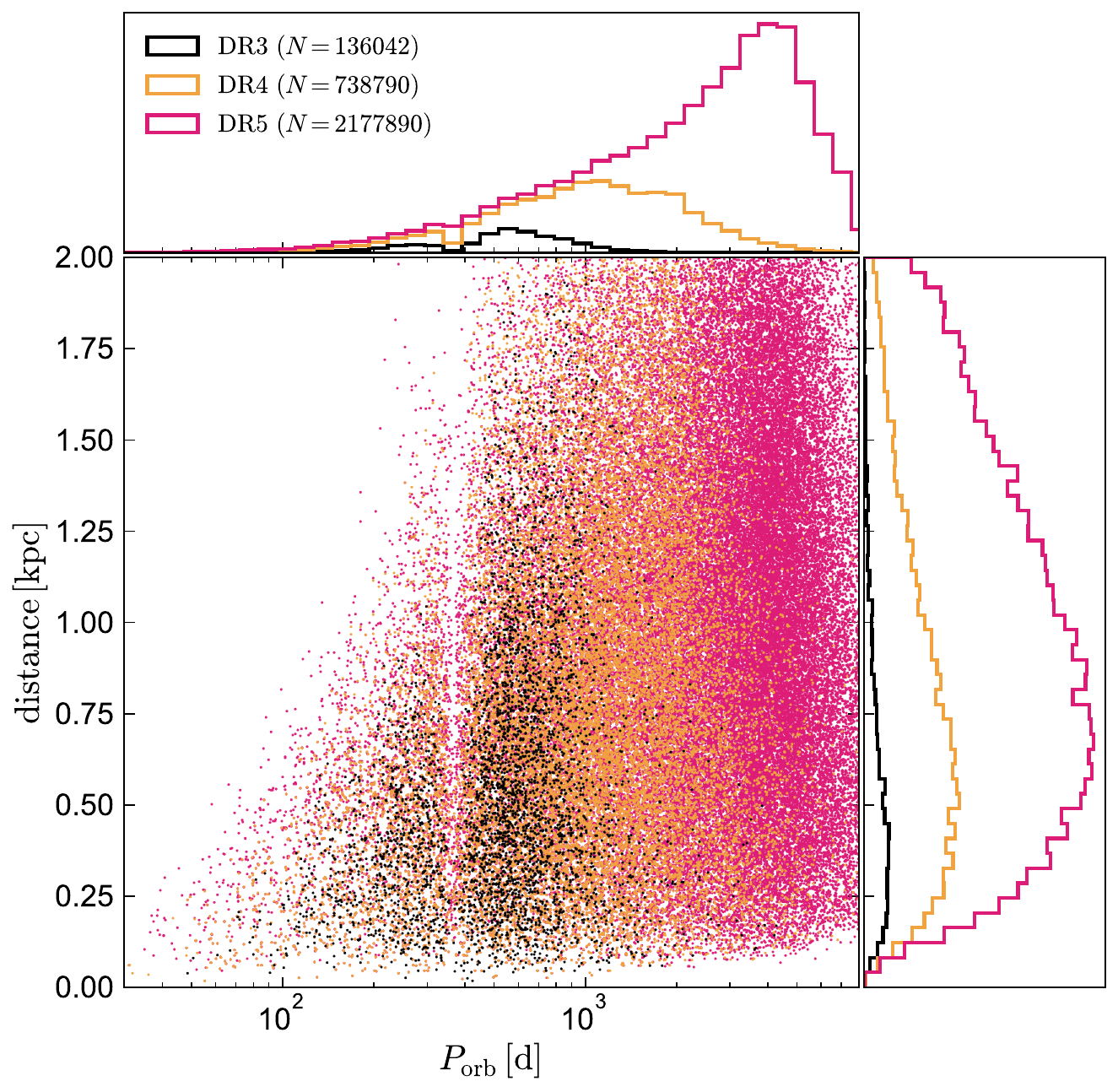}
    \caption{Comparison of predicted binary populations in DR3 (black), DR4 (gold), and DR5 (magenta), assuming exactly the same cuts are employed in future releases as in DR3. 3\% of the full predicted sample is shown. The DR4 and DR5 samples are predicted to be $\sim 5$ and $\sim 12$ times larger than the DR3 sample. This is primarily because these data will be sensitive to binaries with longer periods, and longer-period binaries have larger orbits that can be detected to larger distances.}
    \label{fig:dr4_vs_dr3}
\end{figure}

Figure~\ref{fig:dr4_vs_dr3} compares the predicted yields of the final astrometric catalog in DR3 (black), DR4 (gold), and DR5 (magenta). Here we employ exactly the same quality cuts used in DR3. Our modeling predicts 740,000 million astrometric binaries in DR4, a factor of 5 increase compared to its predictions for DR3. This increase owes both to the longer observing baseline -- leading to a detected period distribution that peaks at $\sim 1600$ days -- and to the increased number of observations per object, which results in smaller parallax uncertainties and better constraints on $a_0$. Because longer-period binaries have larger orbits, they are astrometrically detectable to larger distances, causing the volume within which binaries are detectable to grow rapidly as the observing baseline increases. 

Typical binaries in the DR4 sample are predicted to be somewhat fainter and more distant than in the DR3 sample, with a median distance and apparent magnitude of $d=0.73$ kpc and $G=13.6$, compared to $d=0.50$ kpc and $G=12.9$ in DR3.  An additional improvement anticipated in DR4 is the publication of epoch astrometry for all sources, including those not satisfying Equations~\ref{eq:parallax_cut},~\ref{eq:filters}, and~\ref{eq:filter2}. If we ignore those filters and only require $a_0/\sigma_{a_0} > 5$, our modeling predicts 900,000  binaries in DR4. The improvement compared to the case with no cuts is smaller than in DR3, because the filters primarily remove sources with short periods, for which the search volume is smaller. 

Another significant improvement can be expected in DR5: the same modeling and cuts used in DR3 leads to a predicted 2.2 million orbital solutions in DR5. The true number can be expected to be even larger, because our \texttt{Galaxia} simulation only extends to 2 kpc, and  Figure~\ref{fig:dr4_vs_dr3} suggests that a significant number of binaries detectable in DR5 will be beyond this limit. The period distribution in DR5 is predicted to peak at $P_{\rm orb} \approx 3000$\,d, with a median apparent magnitude of $G=14.4$ and a median distance of $d=0.9$ kpc. 
In the likely event that treatment of systematics improves in future data releases, the expected sample of orbits will be even larger. It could likely also be enlarged by reducing the threshold on $s=a_0/\sigma_{a_0}$ below 5 and by reducing the \texttt{RUWE} threshold below 1.4, as we discuss further in Section~\ref{sec:planets}.

\section{A Monte Carlo selection function}
\label{sec:monte_carlo}
Our modeling makes it possible to ask, for any hypothetical binary with particular properties (sky position, apparent magnitude, period, flux ratio, mass, distance, etc.), whether it would likely have been detectable in a given {\it Gaia} data release: we can simply simulate many realizations of the binary with different orientations, noise draws, etc., fit the mock astrometry for each realization, and calculate what fraction of the mock fits result in a solution that passes all of the quality cuts imposed on the published solutions. The main downside of this approach is that it is slow, and like any Monte Carlo method, it is noisy when the number of simulated realizations is small. For a complementary approach based on analytic approximations of the selection function, see \citet{Casey}.

\begin{figure*}
    \centering
    \includegraphics[width=\textwidth]{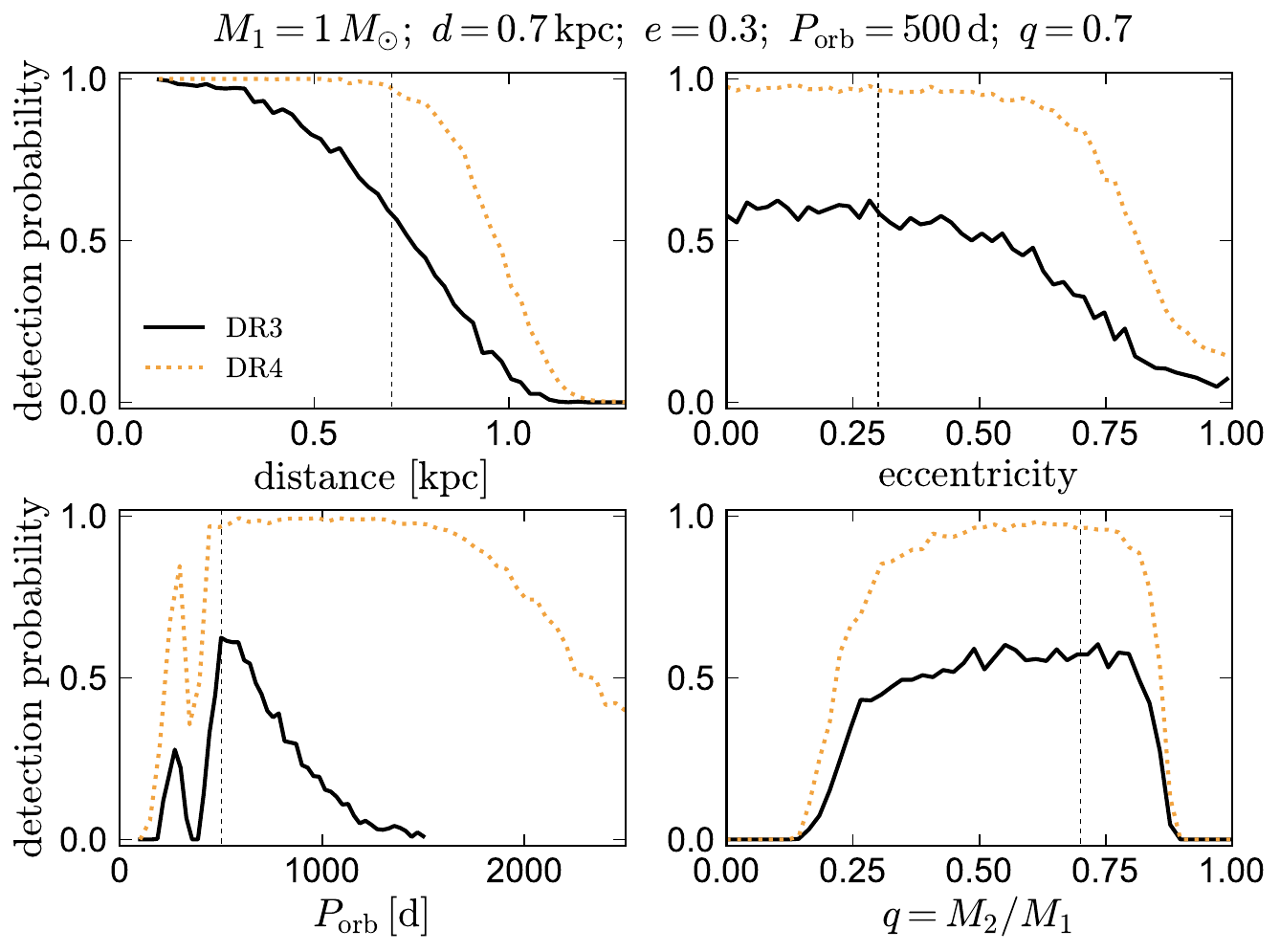}
    \caption{Probability of a binary containing two MS stars being published with an astrometric orbital solution in DR3 (solid black) or DR4 (dotted yellow). For the fiducial case, we take $M_1=1\,M_{\odot}$, $d=0.7$\,kpc, $e=0.3$, $P_{\rm orb}=500$\,d, and $q=0.7$.  From top left to lower right, panels show the effects of varying distance, eccentricity, orbital period, and mass ratio. Other parameters are fixed at their fiducial values, which are marked with dashed vertical lines. For each combination of parameter values, we simulate 500 binaries distributed across the sky, drawn from an exponential disk with scale height $h_z=0.3$\,kpc. We account for position-dependent extinction and {\it Gaia} scanning frequency.   }
    \label{fig:det_prob}
\end{figure*}

\subsection{A main-sequence binary}
Figure~\ref{fig:det_prob} illustrates this approach for a luminous binary. We consider a system containing two MS stars with a $1.0\,M_{\odot}$ primary and mass ratio $q=M_2/M_1=0.7$. In the fiducial case, we consider a distance of 0.7 kpc, a period of 500 d, an eccentricity of 0.3. We simulate 500 realizations of this binary, assigning each one a different sky position and corresponding extinction and scan pattern, and a different random orientation and phase, as quantified by $\Omega$, $\omega$, $i$, and $T_p$. In assigning random positions at a given distance, we assume an underlying stellar population distributed as an exponential disk with scale height $h_z=0.3$ kpc. 

We fit the simulated epoch astrometry for each binary with the astrometric cascade described in Section~\ref{sec:cascade}, and finally calculate what fraction of the simulated binaries lead to accepted orbital solutions that survive all the quality cuts imposed in post-processing. Figure~\ref{fig:det_prob} shows the results of such experiments for a range of distances, eccentricities, orbital periods, and mass ratios. In each panel, we show how the detection probability changes when one property is varied and the rest are held fixed. Black and gold lines show detection probabilities in DR3 and DR4, respectively. 

The detection probability in DR3 falls with distance as expected, but the fall-off is relatively gradual, dropping from 85\% at 0.5 kpc to 60\% at 0.7 kpc, 25\% at 0.9 kpc, and 1\% and 1.1 kpc. The detection probability decreases with increasing eccentricity, but the drop-off only becomes steep at $e>0.6$. The DR3 detection probability is largest at periods of 500-600 days, decreasing at shorter periods due to smaller orbits and at longer periods due to less complete phase coverage, which causes many long-period sources to receive acceleration solutions. Sensitivity to periods of 300-400 days is also severely reduced. Sensitivity varies weakly with $q$ between 0.2 and 0.85 due to the competing effects of a larger orbit at large $q$ and the photocenter moving between the two stars as $q\to 1$ \citep[e.g.][]{Penoyre2020}.  Not surprisingly, all detection probabilities increase in DR4. The increase is most significant at periods longer than 1000 d, which are inaccessible in DR3. 

\subsection{Compact object companions}
\begin{figure*}
    \centering
    \includegraphics[width=\textwidth]{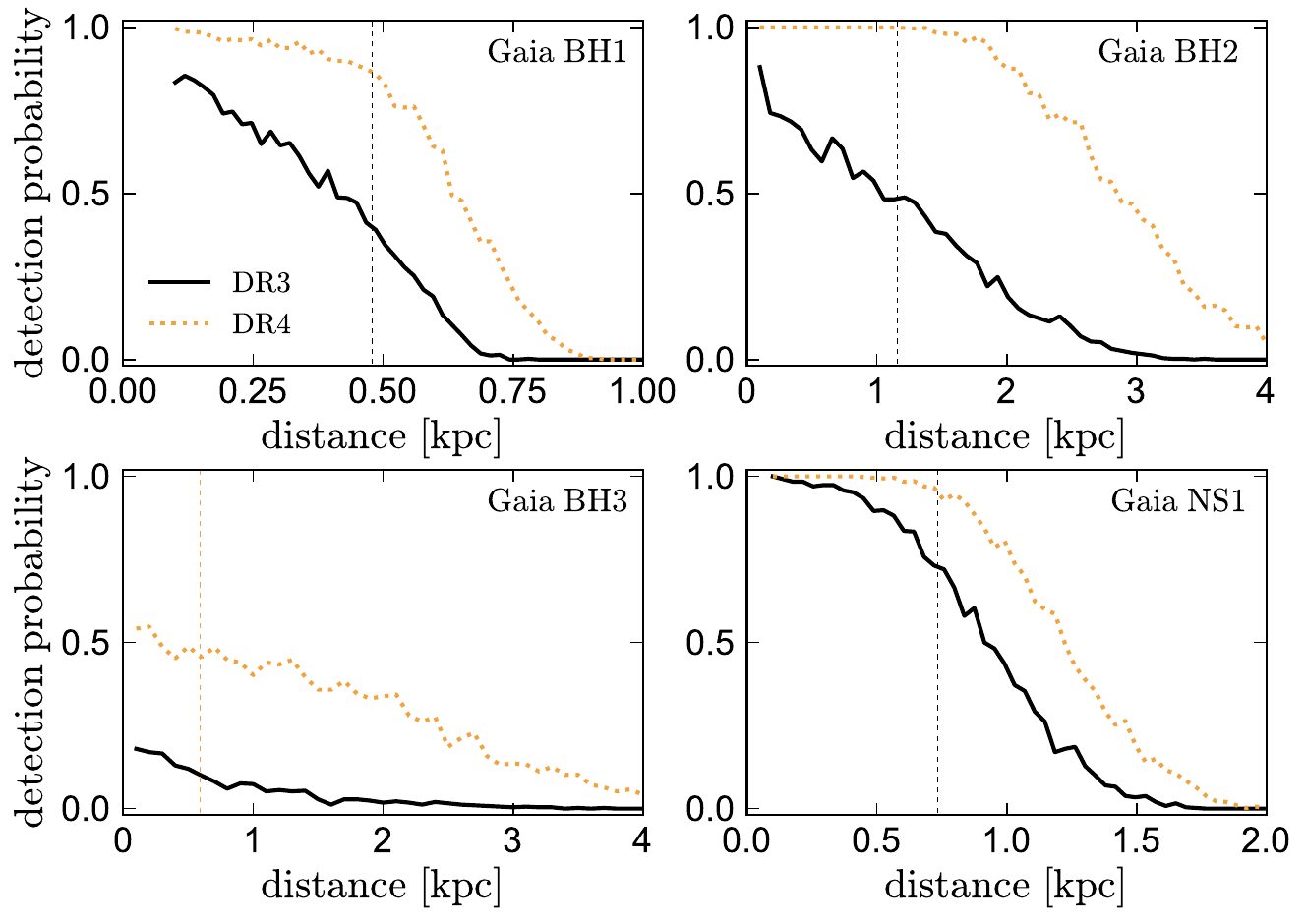}
    \caption{Probability of detecting known BH/NS + luminous star binaries in Gaia DR3 (solid black) and DR4 (dotted gold). At each distance, we simulated 500 realizations of each system, spread across the sky and with different orientations. The properties assumed for each system are listed in Table~\ref{tab:bhs}. Dashed vertical lines show the actual distances to the four systems. }
    \label{fig:det_prob_bhs}
\end{figure*}

Calculations similar to those shown in  Figure~\ref{fig:det_prob} can be carried out for any hypothetical class of binary. As another test case, we consider wide compact object binaries with properties similar to Gaia BH1 \citep{El-Badry2023_bh1, Nagarajan2024}, BH2 \citep{El-Badry2023_bh2}, BH3 \citep{Panuzzo2024}, and Gaia NS1 \citep{El-Badry2024_ns1}. As in the previous case, we generate 500 realizations of each system spread across the sky. The masses, orbital periods, and absolute magnitudes we assume for each system are listed in Table~\ref{tab:bhs}. Figure~\ref{fig:det_prob_bhs} shows detection probabilities as a function of distance. Note that the x-axis varies between panels. 

Both Gaia BH1 and BH2 had a $\approx$50\% probability of being detected and passing all quality cuts to be published in DR3. For BH1, this reflects the fact that the observed parallax significance was only barely high enough to satisfy Equation~\ref{eq:parallax_cut}; for BH2, it is a consequence of the fact that the orbital period is longer than the DR3 observing baseline. In DR4, the observing baseline will be longer and the parallax uncertainties smaller, so the  predicted detection probabilities rise to 80 and $>99\%$. Gaia BH3 has an orbital period much longer than the DR3 observing baseline and thus would have been detectable in DR3 only for the most fortuitous orbital orientations and phases. In DR4, the system's detection probability at its observed distance is 47\%, reflecting the fact that the orbital period still exceeds the observing baseline by more than a factor of 2. Our modeling predicts that Gaia NS1-like binaries would have received an orbital solution for 74\% of orientations and sky positions in DR3, and for 95\% in DR4. 

\begin{table}[h!]
\centering
\caption{Adopted properties of four BH/NS binaries and detection probabilities in {\it Gaia} DR3 and DR4. Detection probabilities as a function of distance are shown in Figure~\ref{fig:det_prob}. }
\label{tab:bhs}
\begin{adjustbox}{max width=\columnwidth}
\begin{tabular}{ccccccccc}
\hline
System & \textbf{$M_{G,0}$} & $P_{\rm orb}$ & $M_\star$ & $M_2$ & $e$ & $d$ & $P_{\rm DR3}$ & $P_{\rm DR4}$ \\
  & [mag] & $[\rm d]$ & $[M_{\odot}]$ & $[M_{\odot}]$ &  & $[\rm kpc]$ & & \\

\hline
Gaia BH1 & 4.50 & 185.4 & 0.93 & 9.3 & 0.43 & 0.48 & 0.39 & 0.86 \\
Gaia BH2 & 1.38 & 1277 & 1.0 & 9.0 & 0.52 & 1.16 & 0.48 & 0.99 \\
Gaia BH3 & 1.72 & 4250 & 0.78 & 33.0 & 0.73 & 0.59 & 0.10 & 0.47 \\
Gaia NS1 & 3.75 & 731 & 0.78 & 1.90 & 0.12 & 0.73 & 0.74 & 0.95 \\
\hline
\end{tabular}
\end{adjustbox}
\end{table}

\begin{figure*}
    \centering
    \includegraphics[width=\textwidth]{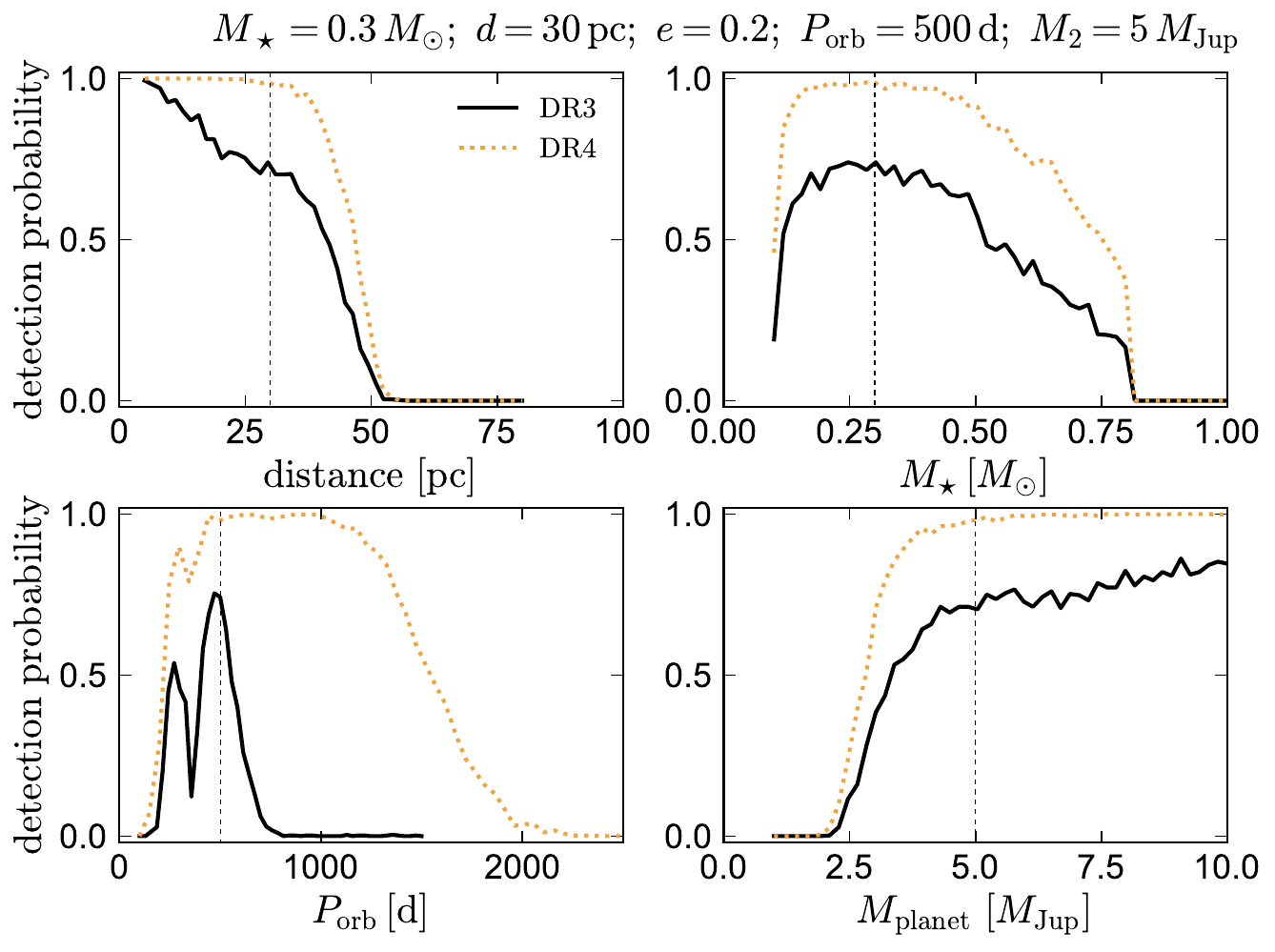}
    \caption{Probability of a giant planet companion published with an astrometric orbital solution in DR3 (solid black) or DR4 (dotted yellow). For the fiducial case, we take $M_\star=0.3\,M_{\odot}$, $d=30$\,pc, $e=0.2$, $P_{\rm orb}=500$\,d, and $M_2=5\,M_{\rm Jup}$.  From top left to lower right, panels show the effects of varying distance, host star mass, orbital period, and planet mass. Other parameters are fixed at their fiducial values, which are marked with dashed vertical lines.  The detection probability is largest for close distances, relatively low-mass stars, periods comparable to the observing baseline, and high-mass planets.  }
    \label{fig:det_prob_planet}
\end{figure*}

\subsection{Giant planets}
\label{sec:planets}
Figure~\ref{fig:det_prob_planet} shows the results of similar calculations for giant planets, which are ultimately expected to be detected by {\it Gaia} in large numbers \citep[e.g.][]{Casertano2008, Sozzetti2014, Perryman2014}. For the fiducial case, we consider a $5\,M_{\rm Jup}$ planet orbiting a $0.3\,M_{\odot}$ star at a distance of 30 pc, with a period of 500 d and and eccentricity of 0.2. We then show the effects of varying distance, stellar mass, orbital period, and planet mass in each panel, and again compare the DR3 and DR4 completeness. We note that the Figure assumes the same conservative quality cuts used in producing the general catalog of astrometric orbits; in reality, looser quality cuts may be adopted for planet candidates \citep[e.g.][]{Holl2023}. 

Figure~\ref{fig:det_prob_planet} suggests that {\it Gaia} DR3 is $>50\%$ complete for the fiducial case within 35 pc, while DR4 will be $>50\%$ complete within 45 pc. The sensitivity varies non-monotonically with $M_\star$ because higher-mass stars wobble less due to a fixed-mass planet but are brighter and thus have more precise astrometry at fixed distance, except at the closest distances where calibration systematics become important. The sharp drop-off visible at $\approx 0.8\,M_{\odot}$ reflects the fact that at $d=30$ pc, stars above this mass are brighter than $G\approx 8$, where {\it Gaia} astrometry becomes significantly less precise (Figure~\ref{fig:noise}). The lower right panel of 
 Figure~\ref{fig:det_prob_planet} shows that {\it Gaia} is much more sensitive to massive planets with $M \gg M_{\rm Jup}$ than to Jupiter-mass or lower-mass planets. 

\begin{figure}
    \centering
    \includegraphics[width=\columnwidth]{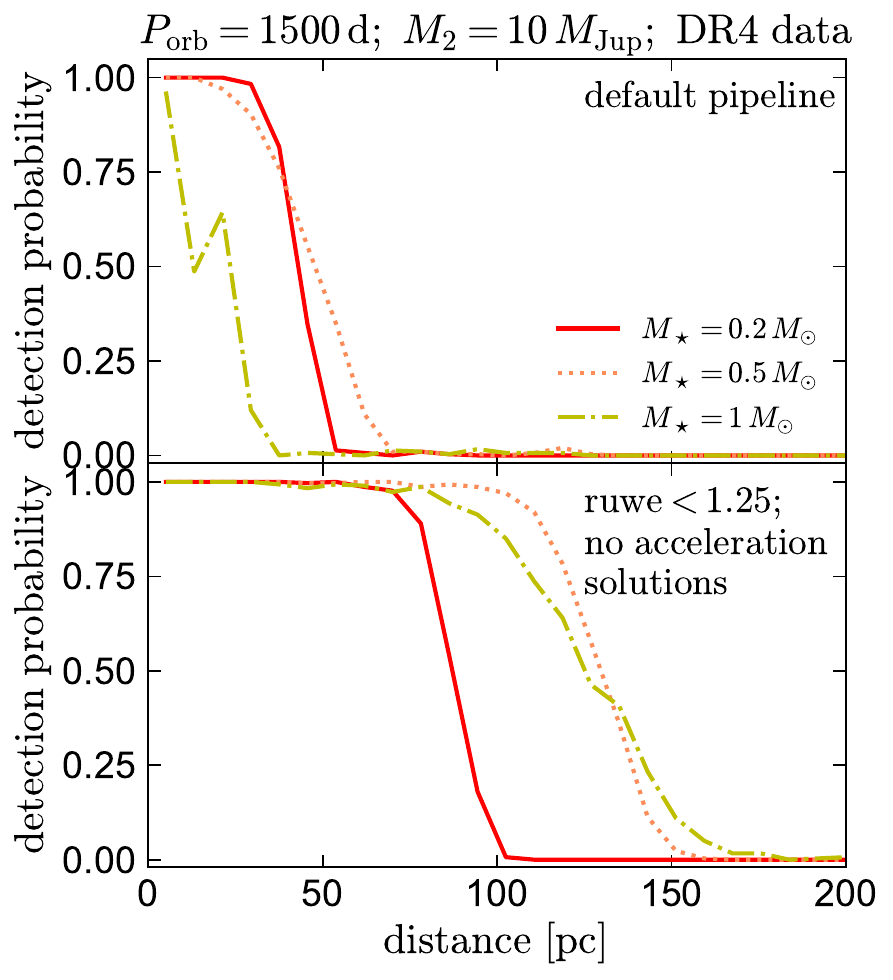}
    \caption{Detection probability for giant planets with $P_{\rm orb} = 1500$ d in DR4. We assume $e=0.2$ and $M_2=10\,M_{\rm Jup}$ and consider three host star masses. Top panel shows predictions for the same cascade of astrometric models and cuts used in producing the DR3 NSS catalog, which lead to detection limits of $\lesssim 50$ pc. Bottom panel shows predictions for a less conservative cascade, where the \texttt{RUWE} threshold is lowered from 1.4 to 1.25 and solutions eligible for an acceleration solution  are not removed from consideration for an orbital solution. This increases the distance limit by a factor of 2 for $0.2\,M_{\odot}$ stars, and by more than a factor of 4 for solar-type stars, enormously increasing the discovery space for giant planets.}  
    \label{fig:planet2}
\end{figure}

In Figure~\ref{fig:planet2}, we explore how the giant planet detection probability varies with distance and host star mass. We simulate DR4 observations of a $10\,M_{\rm Jup}$ planet with $P_{\rm orb} = 1500$ d (near optimal for DR4) and a range of  host star masses. In the top panel, we use the same model cascade and acceptance criteria used in producing the DR3 NSS catalog. In this case, a $10\,M_{\rm Jup}$ companion can only be detected within 20-50 pc, depending on the mass of the luminous star. For $M_\star= 1\,M_\odot$, the non-monotonic decrease in detection probability with increasing distance is a consequence of very bright stars having larger astrometric uncertainties (Figure~\ref{fig:noise}). Trends with $M_\star$ are also nontrivial:  low-mass stars have larger orbits at fixed $M_{\rm planet}$, but they become faint at closer distances. 

We found that beyond the distance limits in the top panel of Figure~\ref{fig:planet2}, giant planets often do not receive orbital solutions because they have UWE $<1.4$, or because they are accepted with acceleration solutions. To explore whether planets at larger distances can be characterized with DR4 data, we tested the effects of using a less conservative astrometric model cascade than was employed in DR3. To this end, we lowered the UWE threshold for consideration with a binary model from 1.4 to 1.25, and we removed the 7- and 9-parameter acceleration models from the astrometric cascade. The lower UWE threshold is motivated by the fact \citep[e.g.][]{Penoyre2022b, Castro-Ginard2024} that \texttt{RUWE} values above 1.25 are not expected due to scatter of well-behaved single-star solutions. We still employed the same quality cuts on the orbital solutions (Equations~\ref{eq:parallax_cut}-\ref{eq:filter2}), so all the accepted orbital solutions are well-constrained. 

The results are shown in the bottom panel of Figure~\ref{fig:planet2}. We find that these relaxed cuts increase the distance limit to which {\it Gaia} can detect a $10\,M_{\rm Jup}$ planet in a 1500-d orbit by a factor of 2-4, corresponding to a 8-64 times larger search volume. These orbits are still fairly well-constrained, with $a_0/\sigma_{a_0} > 5$, allowing the astrometric mass function (Equation~\ref{eq:fm_ast}) to be constrained to within a factor of $\sim 2$. 

The predictions in Figure~\ref{fig:planet2} can be compared to those of \citet{Holl2022}, who forecasted the number of brown dwarfs and giant planets detectable with a 5-year {\it Gaia} mission (comparable to the DR4 observing baseline). They found that a 10\,$M_{\rm Jup}$ planet orbiting a solar-type star could be detected with $>50\%$ probability out to a distance limit of 300 pc, which is somewhat farther than our prediction of 130 pc in the bottom panel of Figure~\ref{fig:planet2}. Similarly, they predict a 50\% completeness limit of 250 pc for a 10\,$M_{\rm Jup}$ planet orbiting a $0.65\,M_{\odot}$ MS star, whereas we find a somewhat closer limit of 150 pc.  We compare to their predictions further in Appendix~\ref{sec:appendix_noncentrality}, concluding that their more optimistic predictions are mainly the result of (a) adopting a lower detectability threshold than the limits that were used for {\it Gaia} orbits published in DR3, and (b) assuming somewhat more optimistic epoch-level astrometric uncertainties than have been achieved thus far. 

The occurrence rate of giant planets in wide orbits is still somewhat uncertain, and we do not attempt to predict the number that will eventually be discovered by {\it Gaia} here. The method we developed can be straightforwardly applied to a simulated planet population using the same strategy we used for binary stars in Section~\ref{sec:mock} \citep[see also][]{Holl2022}. We note that most papers making predictions for {\it Gaia} astrometric exoplanet discoveries in the past assumed a 5-year mission duration, but the actual duration will foreseeable be 11 years. As we showed in Figure~\ref{fig:dr4_vs_dr3}, the expected asymmetric yield from an 11-year mission (DR5) is at least a factor of 3 larger than that of a 5-year mission (DR4), because the search volume increases rapidly with increased sensitivity to long-period orbits.  

Once a planet candidate has been astrometrically identified, the question remains whether it can be unambiguously identified as such. Twin binaries with nearly equal-mass components can always masquerade as planets \citep[e.g.][]{Marcussen2023}, but most of these are overluminous relative to the main sequence, and spectroscopy will reveal all of them to be double-lined. The masses of companions with small but non-negligible flux ratios can also be overestimated from astrometry alone. However, such companions are unlikely to represent a major false-positive for planet searches, because there are few astrophysically plausible companions -- with the possible exception of young brown dwarfs -- with masses and luminosities that will result in planet-like mass functions. 

\section{Summary}
\label{sec:disc}
We have produced an approximate generative model for the sample of astrometric binaries with orbital solutions published in DR3, which appears to capture the most important selection effects of the real catalog. The method can also straightforwardly be applied to future {\it Gaia} data releases. As we will explore in other work, this approach is quite useful for interpreting the contents of observed binary catalogs, performing population inference, and forecasting the contents of future data releases. 

Our key results are as follows:
\begin{enumerate}

\item {\it A public code for generating and fitting epoch astrometry}: We provide Python routines to simulate and fit {\it Gaia} epoch astrometry through the \texttt{gaiamock} code, available \href{https://github.com/kareemelbadry/gaiamock}{here}. The code is suitable for fitting large numbers of sources with a variety of astrometric solutions, as will be possible after {\it Gaia} DR4. It also implements a rejection sampling fitting mode \citep[e.g.][]{Price-Whelan2017} that is likely to perform better at low SNR than the optimization routine used in this work. 

\item {\it Mock catalog}: We use the binary population demographics model of \citet{Moe2017} to populate a synthetic Galaxy with binaries and forward-model their {\it Gaia} observations and astrometric model fitting (Figure~\ref{fig:epoch_astrometry}). We apply the same quality cuts used to construct the {\it Gaia} DR3 astrometric orbit sample to the mock data (Figure~\ref{fig:halbwachs_figure}). This results in a mock catalog whose properties are similar but not identical to the catalog of astrometric binaries published in DR3 (Figures~\ref{fig:catalog_comparison},~\ref{fig:catalog_comparison_sky},~\ref{fig:a0}, and ~\ref{fig:ruwe}). The most significant tension is in the eccentricity distribution, where the mock catalog contains more high-eccentricity orbits.  Varying the properties of the assumed binary population and comparing to the observed catalog is a promising avenue to refine the underlying binary model. 
Such efforts are left to future work.

\item {\it Spurious solutions}: Our forward-modeling predicts a population of spurious orbital solutions with properties similar to the spurious solutions initially selected in DR3, most of which were ultimately filtered out by stringent quality cuts (Figure~\ref{fig:halbwachs_figure}). A large majority of the spurious solutions have one of a few discrete periods related to the {\it Gaia} scanning law and imply implausibly large astrometric mass functions (Figure~\ref{fig:spurious}). Most of the spurious solutions arise from marginally-resolved wide binaries with orbital periods much longer than the {\it Gaia} observational baseline (Appendix~\ref{sec:appendix_resolved}). Many but not all of these systems can be resolved with ground-based high-resolution imaging. This modeling suggests that {\it Gaia} will be significantly less sensitive to astrometric binaries that are in triples, where a distant tertiary will often disturb the astrometry of the inner binary. The post-processing cuts employed in DR3 are effective at removing spurious solutions, but they remove many good solutions as well, particularly at short periods. In future releases, it will likely be possible to retain more of the good solutions by filtering out sources with scan-angle dependent astrometric signals \citep[e.g.][]{Holl2023b}. 

\item {\it Acceleration solutions}: Almost twice as many accelerating solutions as orbital solutions are accepted into the mock catalog. Even more binaries are provisionally accepted with acceleration solutions -- causing them to be removed from consideration for an orbital solution -- but ultimately only published with single-star solutions (Figure~\ref{fig:flowchart}).
The most common true orbital periods for accepted 9- and 7- parameter accelerating solutions in the DR3 mock catalog are $\sim 1000$ and $\sim 1500$ days, respectively  (Figure~\ref{fig:accel}). Many sources for which 9-parameter solutions are accepted could have yielded high-significance orbital solutions but were not fit with orbital solutions because the significance of the acceleration solution was high enough for the source to be removed from further processing in DR3. Acceleration solutions are of limited astrophysical utility -- without a known orbital period or angular separation, only coarse limits can be set on the nature of the companion. Epoch astrometry in DR4 and beyond will allow orbital solutions to be constrained for many of the sources with accelerating solutions in DR3.

\item {\it Predictions for DR4 and DR5}: Our modeling predicts that {\it Gaia} DR4 will yield at least 5 times as many orbital solutions as DR3, even if the same  conservative quality cuts are adopted as in DR3 and the treatment of systematics does not improve. Another factor of $\sim 3$ improvement is expected in DR5 (Figure~\ref{fig:dr4_vs_dr3}). Most of the new solutions will come from binaries with periods longer than 1000 d, which could not be solved in DR3. Such binaries have larger orbits at fixed mass than shorter-period binaries and can thus be astrometrically constrained to larger distance. As a result, the effective search volume for astrometric orbits increases rapidly with the observational baseline. The number of spurious solutions is also expected to decrease significantly in future data releases (Figure~\ref{fig:halbwachs_dr4}). We advocate lowering the \texttt{RUWE} threshold for fitting of orbital solutions in future data releases, as orbits can often be well-constrained even at \texttt{RUWE} $<1.4$. This is particularly true for planets (Figure~\ref{fig:planet2}).

\item {\it Empirical selection function}: Our modeling allows us to construct a Monte Carlo model of the {\it Gaia} selection function: for any hypothetical binary with a given set of observable properties, we can simulate many realizations of the epoch astrometry, varying the binary's orientation, sky position, and noise. We then forward-model the {\it Gaia} orbit fitting and quality cuts and finally calculate what fraction of the simulated realizations would have passed all cuts to be included in the DR3 astrometric binary sample, or the cuts used in future data releases. We provide publically available code to use this selection function.  

Application of this method are shown in Figures~\ref{fig:det_prob}, ~\ref{fig:det_prob_bhs}, ~\ref{fig:det_prob_planet}, and ~\ref{fig:planet2} which consider luminous binaries, compact object companions, and giant planets. One intuitive prediction of this modeling is that {\it Gaia} is significantly less sensitive to high-eccentricity orbits than to low-eccentricity orbits. The bias becomes severe at eccentricities $e \gtrsim 0.6$ (Figure~\ref{fig:det_prob}). Previous work predicted this bias using analytic orbit-averaged models \citep{Penoyre2020, Andrew2022}; our modeling of the {\it Gaia} scanning law and model cascade allows us to quantify its effect on the published binary catalogs.

Our simulations suggest that binaries similar to Gaia BH1, BH2, and BH3 will be detectable to distances of $\approx 0.7$, 3, and 2 kpc, respectively in {\it Gaia} DR4 (Figure~\ref{fig:det_prob_bhs}), suggesting that significant numbers of such binaries will emerge from DR4. Quantitative predictions will be made in future work. The search volume within which most giant planets will be detected will also increase significantly (Figure~\ref{fig:det_prob_planet}).
Massive planets ($M=10\,M_{\rm Jup}$) in few-year periods will be detectable around low-mass and solar-type stars out to $\sim 150$ pc.

\end{enumerate}

\section*{acknowledgments}
We thank the referee for helpful suggestions that improved the manuscript. 
This research was supported by NSF grant AST-2307232 and donations from Isaac Malsky. C.Y.L. acknowledges support from the Harrison and Carnegie Fellowships. HWR acknowledges support from the European Research Council for the ERC Advanced Grant [101054731].

This work has made use of data from the European Space Agency (ESA)
mission {\it Gaia} (\url{https://www.cosmos.esa.int/gaia}), processed by
the {\it Gaia} Data Processing and Analysis Consortium (DPAC,
\url{https://www.cosmos.esa.int/web/gaia/dpac/consortium}). Funding
for the DPAC has been provided by national institutions, in particular
the institutions participating in the {\it Gaia} Multilateral Agreement.

\newpage

\newpage

\bibliographystyle{mnras}

\appendix

\section{Fitting astrometric orbital solutions}
\label{sec:appendix_fitting}

We suppose that a binary is observed at times $t_{i}=\left\{ t_{1},t_{2},\cdots t_{n}\right\}$. Each observation comes with a known scan angle $\psi_i$ and parallax factor $\Pi_{\eta,i}$. The data are the observed along-scan displacements $\eta_i$, and their uncertainties, $\sigma_{\eta,i}$.  For a given guess of $\vec{\mu}_{\rm nonlinear}=\left\{ P,\phi_{0},e\right\} $, we can numerically solve Equation~\ref{eq:kepler} for the eccentric anomalies $E_i$ corresponding to each $t_i$, and thus calculate an array of $X_i$ and $Y_i$ from equation~\ref{eq:XY}. We can then rewrite Equation~\ref{eq:AL} as 

\begin{equation}
\label{eq:linear}
\vec{\eta} = \mathbf{A} \vec{\mu},
\end{equation}
where 


\begin{equation}
    \mathbf{A}=\left[\begin{array}{ccccccccc}
\sin\psi_{1} & t_{1}\sin\psi_{1} & \cos\psi_{1} & t_{1}\cos\psi_{1} & \Pi_{\eta,1} & X_{1}\sin\psi_{1} & Y_{1}\sin\psi_{1} & X_{1}\cos\psi_{1} & Y_{1}\cos\psi_{1}\\
\sin\psi_{2} & t_{2}\sin\psi_{2} & \cos\psi_{2} & t_{2}\cos\psi_{2} & \Pi_{\eta,2} & X_{2}\sin\psi_{2} & Y_{2}\sin\psi_{2} & X_{2}\cos\psi_{2} & Y_{2}\cos\psi_{2}\\
\cdots & \cdots & \cdots & \cdots & \cdots & \cdots & \cdots & \cdots & \cdots\\
\sin\psi_{n} & t_{n}\sin\psi_{n} & \cos\psi_{n} & t_{n}\cos\psi_{n} & \Pi_{\eta,n} & X_{n}\sin\psi_{n} & Y_{n}\sin\psi_{n} & X_{n}\cos\psi_{n} & Y_{n}\cos\psi_{n}
\end{array}\right]
\end{equation} 

is the design matrix, and 

\begin{equation}
    \vec{\eta}=\left[\begin{array}{c}
\eta_{1}\\
\eta_{2}\\
\cdots\\
\eta_{n}
\end{array}\right];\qquad\vec{\mu}=\left[\begin{array}{c}
\Delta\alpha^{*}\\
\mu_{\alpha}^{*}\\
B\\
G\\
\Delta\delta\\
\mu_{\delta}\\
A\\
F\\
\varpi
\end{array}\right]
\end{equation}
The maximum-likelihood solution to Equation~\ref{eq:linear} is

\begin{equation}
    \label{eq:mu}
    \vec{\mu} = \left[ \mathbf{A}^\top \mathbf{C}^{-1} \mathbf{A} \right]^{-1} \left[ \mathbf{A}^\top \mathbf{C}^{-1} \vec{ \eta} \right],
\end{equation}
where $\bf{C}$ is the covariance matrix, 
\begin{align}
    \mathbf{C}=\left[\begin{array}{cccc}
\sigma_{\eta,1}^{2} & 0 & 0 & 0\\
0 & \sigma_{\eta,2}^{2} & \dots & 0\\
0 & \cdots & \cdots & 0\\
0 & 0 & 0 & \sigma_{\eta,n}^{2}
\end{array}\right],
\end{align}
and we assume the individual astrometric measurements are not covariant. 

Solving the orbit is thus reduced to a nonlinear optimization problem in three dimensions. For each guess of $\mu_{\rm nonlinear}$, we can solve Equation ~\ref{eq:mu} for $\mu$, plug the result in Equation~\ref{eq:linear}, and finally evaluate the likelihood, which compares the observed and predicted $\eta_i$ assuming Gaussian uncertainties. The main challenge is in finding find the optimal $\mu_{\rm nonlinear}$, because period aliases can make the posterior multimodal. 

We use adaptive simulated annealing \citep{INGBER1989967}, a global optimization method designed to perform well on bumpy likelihoods. This method has been used previously to fit astrometry \citep{Pourbaix1998}; our implementation borrows from the work of \citet{Iglesias-Marzoa_2015},  who used the method to fit RVs of binaries. We verified using simulations that for a wide variety of orbital periods and configurations the optimizer achieves very similar performance to the \texttt{kepmodel} package \citep{Delisle2022}, which is used in the supplemental code released by \citet{GaiaCollaboration2024} to fit the epoch astrometry for Gaia BH3, but our fitting is $\sim 3$ times faster. This most likely does not reflect any superiority of the optimization algorithm, but is simply a consequence of the fact that the core functions in our implementation are written in C. 

Limited information is available about the details of the fitting algorithm used for {\it Gaia} DR3.
As described by \citet{Halbwachs2023}, the problem was linearized, and the maximum-likelihood solution in 3D was searched for with a nonlinear optimizer. The optimizer was initialized at $e=0$ on a period grid ranging from 10 to $\approx 1600$ d. How dense a grid was searched and the periastron time at which the optimizers were initialized is not specified. Here we proceed under the assumption that the optimization was near-optimal. 

Once the maximum likelihood solution has been identified, the task remains to estimate parameter uncertainties. Properly accounting for uncertainties in both the linear and nonlinear parameters requires returning to twelve dimensions (without analytic marginalization over the linear parameters). We approximate the covariance matrix of the astrometric parameters as 

\begin{equation}
    \mathbf{C}_{\rm ast} = (\mathbf{J}^\top \mathbf{J})^{-1},
\end{equation}
where $\mathbf{J}$ is the Jacobian matrix, defined as 
\begin{equation}
    \mathbf{J}_{ij} = \frac{\partial r_i}{\partial \theta_j}. 
\end{equation}
Here $r_{i}=\left(\eta_{{\rm pred},i}-\eta_{i}\right)/\sigma_{\eta,i}$ is the uncertainty-scaled residual of the $i$th datapoint, and $\theta_j$ is the $j$th astrometric parameter; we evaluate these derivatives numerically using a step size of $10^{-8}$ in all parameters. Finally, we take the uncertainties of the astrometric parameters as the square root of the diagonal elements of the covariance matrix, and we transform Thiele-Innes elements into Campbell elements following \citet{Halbwachs2023}. 

We also experimented with using a Markov Chain Monte Carlo approach. We found that for well-constrained solutions that pass all the cuts imposed in DR3, this yields uncertainties in good agreement with those inferred from the linear analysis. We opted to use the linear approach because it more closely matches the method used by \citet{Halbwachs2023} and because it is faster, particularly for slow-to-converge solutions with periods longer than the observing baseline. 



\section{Effects of IPD resolved binary cuts}
\label{sec:appendix_ipd}

\begin{figure}
    \centering
    \includegraphics[width=0.6\columnwidth]{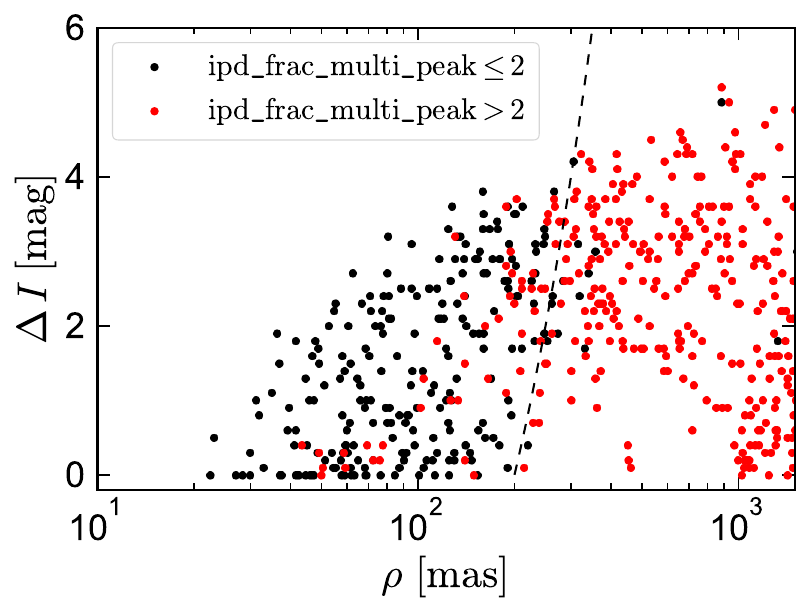}
    \caption{Magnitude difference vs. angular separation for wide pairs resolved by \citet{Tokovinin2023} with speckle imaging. Red points show pairs in which the primary has \texttt{ipd\_frac\_multi\_peak} $>2$, and thus would not have been eligible for an astrometric binary solution. Black points show pairs with \texttt{ipd\_frac\_multi\_peak} $\leq2$, which would have been eligible for a solution. Dashed black line shows $\Delta I\,\left[{\rm mag}\right]=\frac{1}{25}\left(\frac{\rho}{{\rm mas}}-200\right)$.}
    \label{fig:ipd_fig}
\end{figure}

As described by \citet{Halbwachs2023}, one of the first steps in  selecting astrometric binary candidates in {\it Gaia} DR3 was the removal of sources with \texttt{ipd\_frac\_multi\_peak} $>2$ of \texttt{ipd\_gof\_harmonic\_amplitude} $>0.1$. This cut was designed to remove marginally resolved sources, whose astrometry is likely to be problematic if not subject to special processing. 

In order to assess the likely effects of these cuts on our simulated binary sample, we investigated the  \texttt{ipd\_frac\_multi\_peak} and \texttt{ipd\_gof\_harmonic\_amplitude} values of a sample of close pairs resolved by \citet{Tokovinin2023} with speckle imaging in the $I$-band. As shown in Figure~\ref{fig:ipd_fig}, we found that most pairs with $\rho \gtrsim 200$ mas have \texttt{ipd\_frac\_multi\_peak} $>2$, meaning that two separate peaks were detected in at least 2\% of CCD observations. Most pairs closer than 200 mas have \texttt{ipd\_frac\_multi\_peak} = 0, and pairs with larger flux ratios have  \texttt{ipd\_frac\_multi\_peak} $< 2$ out to slightly larger separations. We found that few sources satisfy \texttt{ipd\_gof\_harmonic\_amplitude} $>0.1$ at any separation, so this cut has little additional effect on the selection. 

The dashed black line in Figure~\ref{fig:ipd_fig} shows an empirical boundary,

\begin{equation}
    \label{eq:delta_G}
    \Delta I\,\left[{\rm mag}\right]=\frac{1}{25}\left(\frac{\rho}{{\rm mas}}-200\right).
\end{equation}
Most sources below this boundary will be discarded and will not be fit with astrometric binary models. In our simulations, we apply the same cut in the $G$-band. Although the $G$- and $I$-bands are not identical, the dependence on $\Delta I$ in Figure~\ref{fig:ipd_fig} is relatively weak, so we expect this to be a servicable approximation. 

\section{Parallax uncertainties of 5-parameter solutions}
\label{sec:appendix_singlestar}

\begin{figure}
    \centering
    \includegraphics[width=\columnwidth]{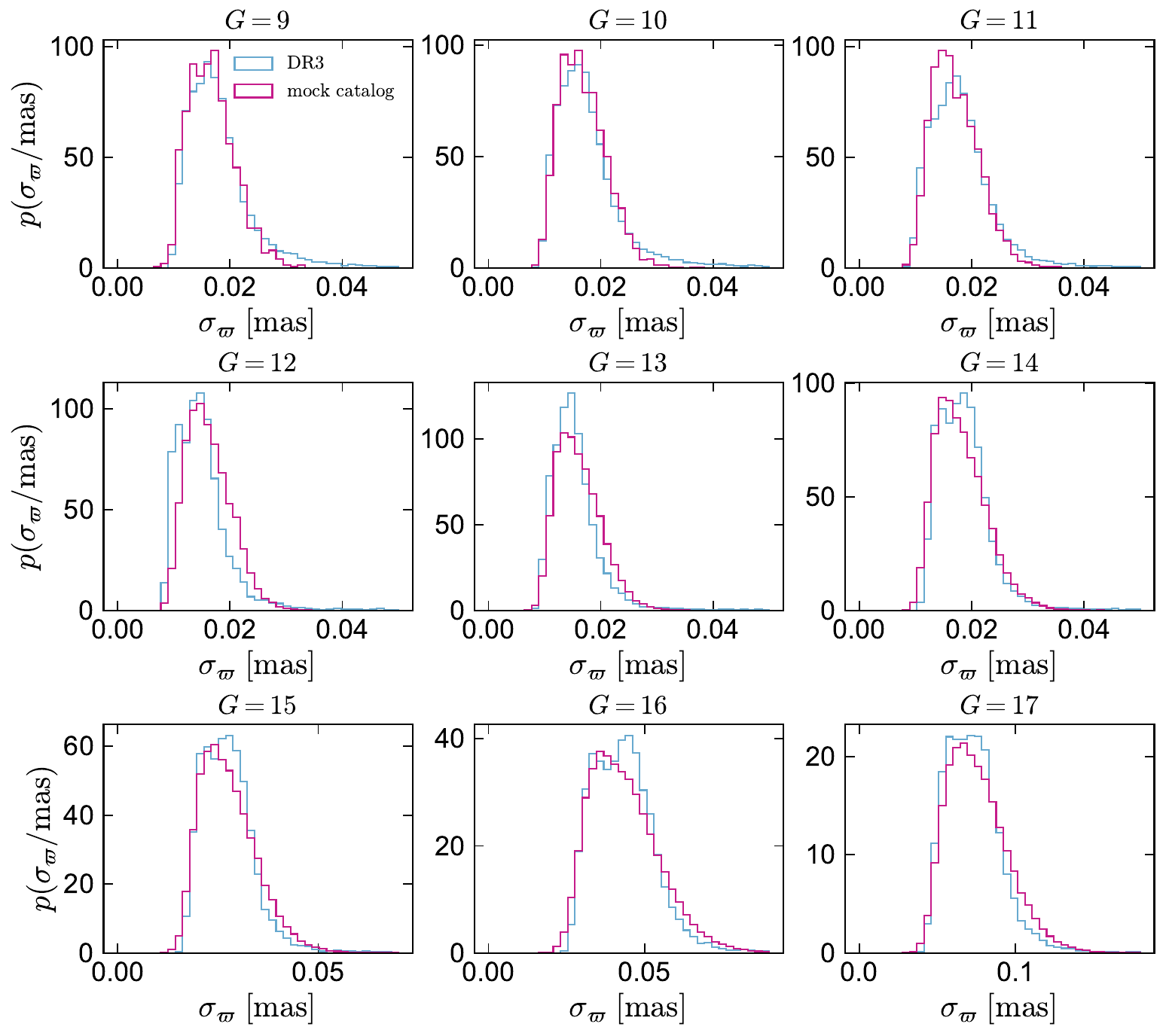}
    \caption{Parallax uncertainty distributions for single-star solutions with \texttt{ruwe} $< 1.4$ in DR3 (blue) and in the mock catalog (violet). Each panel shows a 0.5 mag-wide bin apparent magnitude bin centered on the value listed in the title. The generally good agreement between the observed and simulated distributions implies that our simulated epoch-level astrometric uncertainties are reasonably accurate, and in particular that measurements from 9 CCDs with a single FOV transit are nearly independent.  }
    \label{fig:singlestars}
\end{figure}

To validate the per-FOV transit uncertainties $\sigma_{\eta}$ used in our modeling (Figure~\ref{fig:noise}), we generated epoch astrometry for a population of simulated single stars, again using \texttt{Galaxia} to model the underlying stellar population. We simulate 1\% of the sources with $G<19$ within 2 kpc, again fitting the cascade of astrometric models used in DR3. All sources ultimately receive 5-parameter solutions. In Figure~\ref{fig:singlestars}, we compare the $\sigma_\varpi$ distributions for the simulated 5-parameter solutions of all sources with $\rm UWE < 1.4$ to those in DR3 with \texttt{ruwe} $<1.4$. Each panel shows a 0.5-mag wide apparent magnitude bin (e.g., the $G=9$ bin includes sources with $G=8.75-9.25$). Overall, the simulated sources reproduce the observed $\sigma_\varpi$ distributions quite well. This implies that our assumed uncertainties are realistic, at least on average, since for a fixed scanning law, the astrometric parameter uncertainties scale linearly with the epoch-level astrometric uncertainties. 

\section{Comparison to previous work}
\label{sec:appendix_noncentrality}

In forecasting exoplanet discoveries with {\it Gaia}, \citet{Perryman2014}  quantified detectable via $\Delta \chi^2$, the difference between the $\chi^2$ of the best-fit 5 parameter solution and that of the correct 12-parameter solution. An advantage of quantifying detectability in terms of $\Delta \chi^2$ is that it is inexpensive to calculate, since it does not require blind fitting of the orbital solution. Modeling a five-year {\it Gaia} mission, they proposed a threshold of $\Delta \chi^2 \gtrsim 50$ as representing a robust detection, and $\Delta \chi^2 \gtrsim 100$ as yielding orbital parameter constraints with uncertainties of 10\% or better. These same thresholds were used by \citet{Holl2022} in forecasting brown dwarf discoveries. 

\begin{figure}
    \centering
    \includegraphics[width=0.5\linewidth]{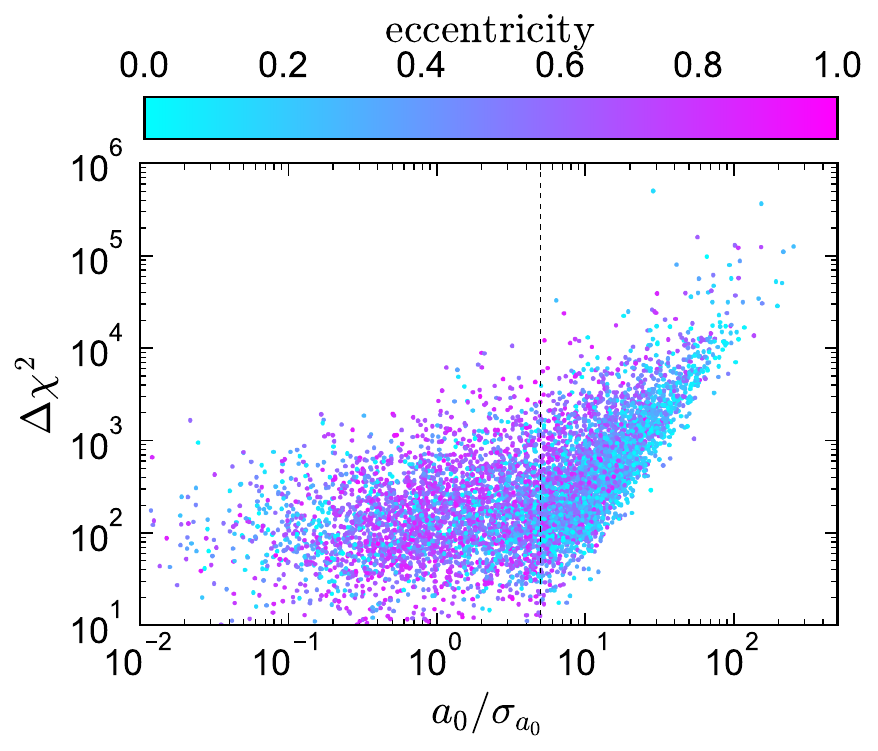}
    \caption{$\chi^2$ difference between 5-parameter single-star and 12-parameter orbital solutions for simulated binaries, vs. signal-to-noise ratio of the photocenter semi-major axis. Dashed vertical line shows $a_0/\sigma_{a_0} = 5$, the minimum threshold adopted in DR3. This threshold corresponds roughly to a cut of $\Delta \chi^2 \gtrsim 160$, but there is  considerable dispersion in $\Delta \chi^2$ at fixed $a_0/\sigma_{a_0}$. Points are colored by their true eccentricity. Some binaries with $\Delta \chi^2$ up to 1000 -- mainly systems with high eccentricity -- have poorly constrained $a_0$ despite having relatively large $\Delta \chi^2$. Systems with lower eccentricities generally have higher $a_0/\sigma_{a_0}$ at fixed $\Delta \chi^2$. }
    \label{fig:delta_chi2}
\end{figure}

To compare to these works, we calculated $\Delta \chi^2$ values for a random subset of $10^6$ simulated binaries used in constructing the mock catalog. Figure~\ref{fig:delta_chi2} shows the distribution of $\Delta \chi^2$ and $s_{\rm orb} =a_0/\sigma_{a_0}$ values for systems that are fit with orbital solutions (i.e., those with $\rm UWE > 1.4$ that are not accepted with acceleration solutions).  At $s_{\rm orb} \gtrsim 5$, there is a clear correlation between the two quantities, but there is also significant scatter. Orbits with low eccentricity typically have the highest $s_{\rm orb}$ at fixed $\Delta \chi^2$. The median $\Delta \chi^2$ at $s_{\rm orb}=5$ is 160, somewhat higher than the thresholds used by \citet{Perryman2014} and \citet{Holl2022}. There is also a significant population of systems with $\Delta \chi^2$ up to $10^3$ and $s_{\rm orb} \ll 5$. These systems all have best-fit eccentricities near unity, and most have true eccentricities $\gtrsim 0.8$. A handful of low-eccentricity binaries with spurious orbital solutions also enter the sample. The astrometric uncertainties blow up in the limit of $e\to 1$, and as a result, many high-eccentricity binaries will not sufficiently well-constrained orbits to enter the DR3 binary sample. 

\section{Astrometry for marginally resolved sources}

\begin{figure*}
    \centering
    \includegraphics[width=\textwidth]{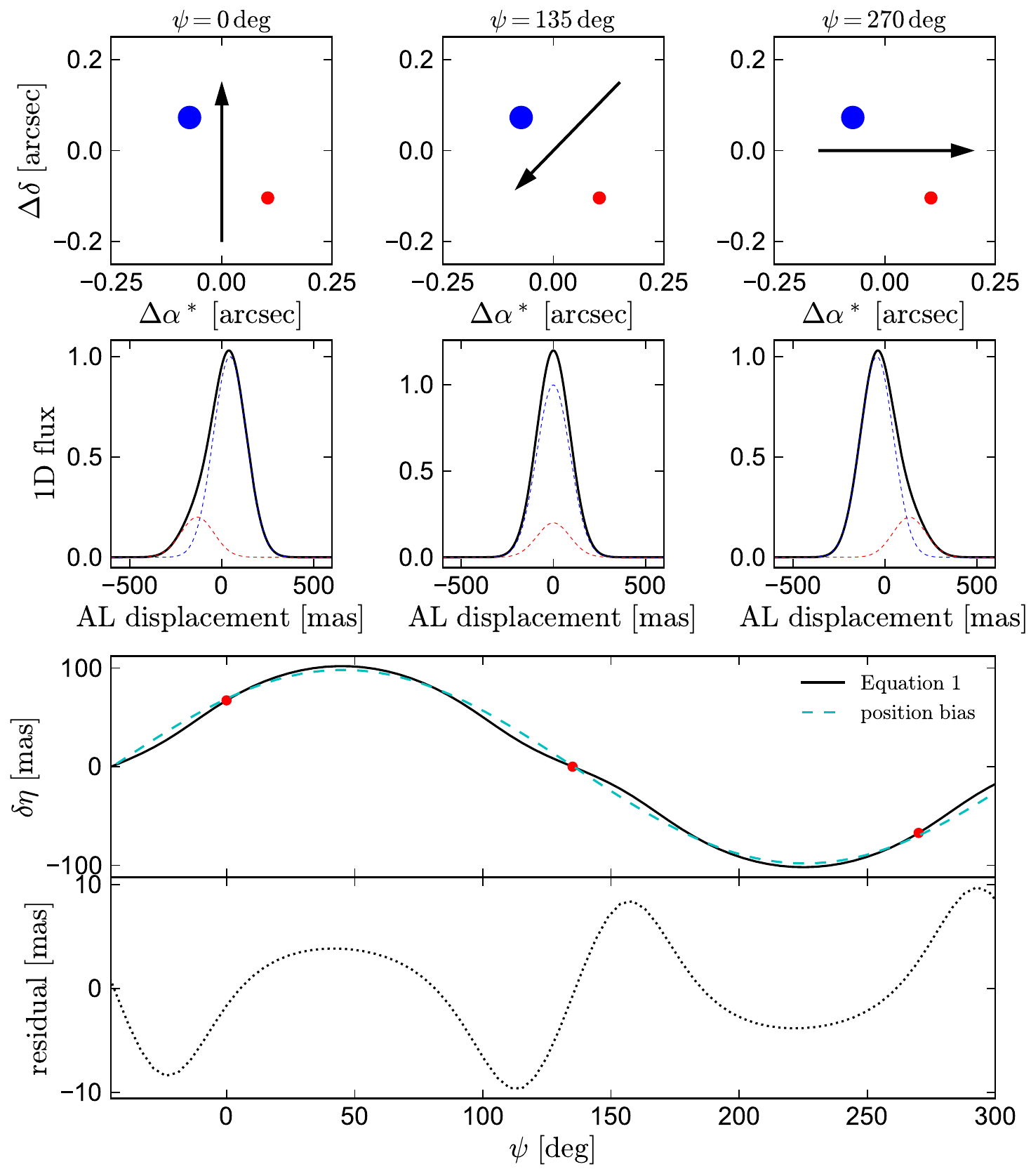}
    \caption{Illustration of how marginally resolved wide binaries lead to spurious astrometric orbits. Top panels show a pair separated by 0.25 arcsec, with the scan angle indicated by a black arrow. 2nd row shows the total flux integrated over the across-scan direction. Displacement is relative to the pair's center of mass, which is at the origin. 3rd panel shows $\delta \eta$ as predicted by Equation~\ref{eq:bias}, with red points corresponding to the panels above. Dashed cyan line shows the best-fit sinusoid, which is absorbed as a positional bias. The residual with respect to this sinusoid (bottom panel) manifests as a spurious astrometric wobble. }
    \label{fig:schematic}
\end{figure*}

Figure~\ref{fig:schematic} illustrates the effects of marginally resolved binaries on {\it Gaia} astrometry. We consider a MS+MS binary with a mass ratio $q=0.7$, a $G-$band flux ratio $f=0.2$, and a projected separation $\rho=250$\,mas. The top two rows show the positions of the two stars relative to the center of mass, and the expected total 1D flux profile as a function of AL coordinate, all for three different scan angles. The peak of the total flux profile shifts by $\sim$100 mas across observations. The third row shows $\delta \eta$ predicted by Equation~\ref{eq:bias} as a function of scan angle. The total variation has an amplitude of about 100 mas, but most of this simply reflects position bias: i.e., the fact that the photocenter and barycenter are not in the same place. Position bias will be absorbed into the best fit solution, so we fit the black line in the 3rd panel with a function $\delta\eta=c_{1}\sin\psi+c_{2}\cos\psi$, where $c_1$ and $c_2$ are free parameters that correspond to the position bias in $\Delta \alpha^*$ and $\Delta \delta$ (Equation~\ref{eq:eta_5par}). The best-fit position bias is shown with a dashed cyan line, and the residuals are shown in the 4th panel. It is these residuals that lead to spurious astrometric wobble. More examples of such residuals can be found in Figures 20 and 21 of \citet{Holl2023b}.

\label{sec:appendix_resolved}

\end{document}